\documentclass[twocolumn,amsmath,amssymb,aps,pra,superscriptaddress,showpacs,10pt]{revtex4-1}

\usepackage{graphicx}
\usepackage{grffile} 

\usepackage{bbm}
\usepackage{color}
\usepackage[utf8]{inputenc}
\usepackage[OT1]{fontenc}

\newcommand{\ket}[1]{{\mid\!#1\,\rangle}}

\newcommand{\braket}[2]{{\langle\, #1\! \mid\! #2 \,\rangle}}

\newcommand{\ie}{i.e., }
\newcommand{\eg}{e.g., }
\newcommand{\fig}[1]{Fig.~#1}
\newcommand{\eq}[1]{Eq.~(#1)}
\newcommand{\tab}[1]{Table #1}
\newcommand{\sect}[1]{Sec.~#1}
\newcommand{\cf}{cf.\ }

\newcommand{\nee}{N_{ee}}
\newcommand{\neel}[1]{N^{[#1]}_{ee}}
\newcommand{\Eint}{E_{\text{int}}}

\begin{document}
\title{Spectra and ground states of one- and two-dimensional laser-driven lattices of ultracold Rydberg atoms}

\pacs{67.85.-d, 32.80.Rm, 32.80.Ee, 37.10.Jk}

 \author{Wolfgang Zeller}
 \affiliation{Zentrum f\"ur Optische Quantentechnologien, Universit\"at Hamburg, Luruper Chaussee 149, 22761 Hamburg, Germany}
 
 \author{Michael Mayle}
 \affiliation{JILA, University of Colorado and National Institute of Standards and Technology, Boulder, Colorado 80309-0440, USA.}
 
 \author{Thorsten Bonato}
 \affiliation{Institute of Computer Science, University of Heidelberg, Im Neuenheimer Feld 368, 69120 Heidelberg, Germany}

 \author{Gerhard Reinelt}
 \affiliation{Institute of Computer Science, University of Heidelberg, Im Neuenheimer Feld 368, 69120 Heidelberg, Germany}

 \author{Peter Schmelcher}
 \email[]{peter.schmelcher@physnet.uni-hamburg.de}
 \affiliation{Zentrum f\"ur Optische Quantentechnologien, Universit\"at Hamburg, Luruper Chaussee 149, 22761 Hamburg, Germany}

\date{\today}

\begin{abstract}\label{txt:abstract}
We investigate static properties of laser-driven, ultracold Rydberg atoms confined to one- and two-dimensional uniform lattices in the limit of vanishing laser coupling. The spectral structure of square lattices is compared to those of linear chains and similarities as well as differences are pointed out. Furthermore, we employ a method based on elements of graph theory to numerically determine the laser detuning-dependent ground states of various lattice geometries. Ground states for chains as well as square and rectangular lattices are provided and discussed. 
\end{abstract}

\maketitle
\section{Introduction} 
\label{sec:introduction}
Recent years have seen an increasing interest in the coherent control of ultracold Rydberg states, \ie states with large principle quantum number $n$ with energies just below the ionization limit. The vast displacement of the valence electron leads to a large polarizability and therefore makes Rydberg atoms very susceptible to external fields \cite{0521385318}. Furthermore, these polarizabilities can lead to very strong Rydberg-Rydberg interactions of dipole-dipole or van-der-Waals character. These interactions can be tuned by applying external fields as well as by the choice of the atomic state. The so-called blockade mechanism resulting from the strong mutual interaction is an essential building block in many works \cite{RevModPhys.82.2313}, as it is in the present one. In a nutshell, the Rydberg blockade mechanism means that the strong Rydberg-Rydberg interaction induces a level shift of many-body states with more than one Rydberg atom present such that the excitation of an atom in the vicinity of a Rydberg atom is shifted off-resonant and is therefore prevented (blocked). The dipole-blockade mechanism has been predicted theoretically in works addressing quantum information processing for the implementations of quantum logic gates \cite{PhysRevLett.85.2208, PhysRevLett.87.037901}. Few years later, various groups succeeded to demonstrate the blockade mechanism also experimentally in ensembles of rubidium atoms  \cite{PhysRevLett.93.063001, PhysRevLett.93.163001, PhysRevLett.95.253002} and cesium atoms \cite{PhysRevLett.99.073002}. More recently, two experiments demonstrated the Rydberg blockade effect for two individual atoms with a distance of several micrometers \cite{Urban2009, Gaetan2009} and subsequently the realization of a {\sc cnot} gate and the generation of entanglement \cite{PhysRevLett.104.010502,PhysRevLett.104.010503}. In \cite{PhysRevLett.99.163601, Gaetan2009} the associated enhanced collective Rabi frequency has been experimentally verified.

The ability to precisely control the trapping potential of Rydberg atoms by means of magnetic \cite{mayle:053410} and optical \cite{PhysRevLett.107.263001} traps in combination with the long-range mutual interaction put a lot of attention of theoretical works to Rydberg atoms in lattices. For instance, in \cite{PhysRevA.79.043419,PhysRevLett.103.185302, PhysRevA.81.023604} a one-dimensional lattice with periodic boundary conditions, \ie a ring lattice, is considered. In the regime of a weak laser coupling with respect to the interaction strength of neighboring Rydberg atoms, the time evolution of Rydberg densities and correlations has been analyzed, whereas in the regime of a strong laser coupling an analytical approach lead to collective fermionic excitations. In a recent work we have presented spectral properties of finite laser-driven lattices with uniform as well as variable spacings \cite{Tezak2010a}. This knowledge can be exploited to determine the strength of the Rydberg-Rydberg interaction by means of the excitation dynamics \cite{Mayle2011}.

Several works investigate the ground states of Rydberg lattices since they are premier candidates for an experimental realization. In \cite{PhysRevLett.104.043002} a method to selectively excite crystalline structures of Rydberg atoms by employing a chirped laser pulse is proposed. The ground state of a one-dimensional lattice has been analyzed in \cite{PhysRevLett.105.230403} by treating the influence of a finite laser coupling perturbatively. The resulting phase diagram of the system, which is also investigated in \cite{PhysRevB.84.085434} in more detail, shows a quantum melting of the crystalline phases. In \cite{PhysRevLett.106.025301} a parameter regime has been identified which allows an analytical solution for the entangled many-body ground state of a one-dimensional Rydberg gas. Considering also the possibility of the Rydberg atom to spontaneously decay back into the ground state, in \cite{PhysRevA.84.031402} the occurrence of an antiferromagnetic phase transition has been demonstrated theoretically, while in \cite{PhysRevLett.108.023602} collective quantum jumps between states of low and high Rydberg population are predicted. There are also first theoretical works on ground states of two dimensional lattices, which indicate phase transitions from an ordered to a disordered phase \cite{Ji2011}. In the regime of strong laser driving, two-dimensional Rydberg gases have also been shown to be good candidates for the creation of collective many-body states that might act as deterministic single-photon sources \cite{Laycock2011}.

The present work expands our previous investigations of laser-driven lattices of ultracold Rydberg atoms to two dimensions. Our particular focus is on the determination of the laser detuning-dependent ground states for different lattice geometries. The knowledge of these ground states is a prerequisite for the selective creation of ordered many-body states in ensembles of Rydberg atoms, as was proposed in \cite{PhysRevLett.104.043002} for the one-dimensional case. Moreover, similarly to [22], the ground states at hand can serve as a first step for perturbatively analyzing the phase diagram in two dimensions. Because of the enormous state space already encountered for relatively small lattice dimensions, sophisticated mathematical tools are necessary to handle the ground state determination even in the presumably simple case of vanishing laser coupling. Here, we show that the concept of maximum cuts -- a technique in graph theory for computing the ground state of spin glasses \cite{9783527404063} -- can be employed for the considered laser-driven lattices of ultracold Rydberg atoms.

The basic model, the associated Hamiltonian and its symmetries are provided in \sect{\ref{sec:model}}. Subsequently, in \sect{\ref{sec:spectra}} we compare the spectra of chains and square lattices in the limit of vanishing laser coupling. Section \ref{sec:ground_states}, which is addressing the laser detuning-dependent ground states, is divided as follows: After defining the ground state problem, we show in \sect{\ref{sec:gs:max-cut}} that it is equivalent to the so-called maximum cut problem. Section \ref{sec:gs:seq} is dedicated to the numerical results of the maximum cut problem, wherein we provide the ground states for some chains, square lattices and an example of rectangular lattices. We conclude with a brief summary (\sect{\ref{sec:summary}}) and an appendix (\sect{\ref{sec:app:appendix}}), which contains a more detailed calculation concerning the Ising-type form of the Hamiltonian used in \sect{\ref{sec:ground_states}}.

\section{The Model} 
\label{sec:model}

Throughout this work we consider the same model as in \cite{Tezak2010a,Mayle2011}: An ultracold gas is trapped in a finite, regular one- or two-dimensional optical lattice with in total $N$ sites, each containing $N_0$ atoms. An additional external laser field couples the ground state to a Rydberg level by means of a two-photon transition with a Rabi frequency of $\Omega_0$. If the excitation lasers are far detuned from the intermediate level, it can be adiabatically eliminated as demonstrated in \cite{Mayle2010a}. We also allow for a small detuning of the two-photon transition with respect to the Rydberg level. Because of the adiabatic elimination of the intermediate level, the single-atom model so far reduces to a two-level system. When assuming in addition that the blockade radius is larger than the spatial extension of the individual sites, each site $k$ can again be described as a two-level system with a collective ground state $\ket{g}_k$ and an excited state $\ket{e}_k$ which is characterized by a symmetrically shared Rydberg excitation. The coupling between the levels of this so-called \emph{superatom} is thus given by the enhanced collective Rabi frequency $\Omega=\sqrt{N_0}\Omega_0$.

\subsection{Hamiltonian}

Taking into account the Rydberg interaction between different sites, the final model Hamiltonian of the system reads in the basis of the superatom states $\{\ket{g}_k,\ket{e}_k\}$ and after applying the rotating wave approximation
\begin{equation}
H_{\text{mod}} = \frac{\Delta}{2} \sum_{k=1}^{N}\sigma^{(k)}_z + \frac{\Omega}{2} \sum_{k=1}^{N} \sigma^{(k)}_x + \sum_{k=1}^{N-1} \sum_{j=k+1}^{N} V_{k,j} n^{(k)}_e n^{(j)}_e.
\label{eq:mod:gen_ham}
\end{equation}
Here $\sigma^{(k)}_x, \sigma^{(k)}_z$ are the Pauli matrices and $n^{(k)}_e = \frac12 [ \sigma^{(k)}_z + \mathbbm{1}]$ is the excitation number operator, where the label $k$ denotes the site index.
Equation (\ref{eq:mod:gen_ham}) can also be interpreted as the Hamiltonian of an interacting spin chain subject to two external orthogonal magnetic fields.

The first term $H_{\text{det}} = \frac{\Delta}{2} \sum_{k=1}^{N} \sigma^{(k)}_z$ contains the laser detuning $\Delta$ which represents the energy spacing between the ground state $\ket{g}_k$ and the excited state $\ket{e}_k$ of the superatom in the rotated frame of reference. 
The second contribution $H_{\text{coup}} = \frac{\Omega}{2} \sum_{k=1}^{N} \sigma^{(k)}_x$ is off-diagonal and couples the two levels $\ket{e}_k$ and $\ket{g}_k$ via the collective Rabi frequency $\Omega$.
The remaining part $H_{\text{int}} = \sum_{k=1}^{N-1} \sum_{j=k+1}^{N} V_{k,j} n^{(k)}_e n^{(j)}_e$ represents a two-body interaction between Rydberg excitations. More precisely, we consider a long-range van der Waals interaction between Rydberg excitations,
\begin{equation}
 V_{k,j}=\frac{C_6}{|\mathbf{r}_k-\mathbf{r}_j|^6},
\end{equation}
where $\mathbf{r}_k$ is the position of the $k$-th site in the lattice, \ie the interaction depends on the spatial separation between the sites. In this work we assume a repulsive interaction,  $V_{k,j} >0$, which is common for Rydberg atoms in their $ns$ state \cite{0953-4075-38-2-021}.
In the configuration of a regular 1D chain with lattice constant $a$, the distance between two sites $k$ and $j$ is simply $|j-k|a$. Thus, the interaction strength reads
\begin{equation}
V_{k,j} = \frac{1}{|j-k|^6} V_1,
\label{eq:mod:vkj_chain}
\end{equation}
where we defined $V_1:={C_6}/{a^6}$.

For the case of regular squares and rectangular lattices with spacing $a$, the $N$ sites are assumed to be arranged in $M$ columns and $N/M\leq M$ rows such that the site index $k$ increases as follows:
\begin{equation}
\left[\begin{matrix}
 1 & 2 & \cdots & {M} \\
 {M+1} & {M+2} & \cdots & {2M} \\
 \vdots & \vdots & \ddots & \vdots \\
 {N-M+1} & {N-M+2} & \cdots & N
\end{matrix}\right].
\label{eq:mod:lattice_row_column}
\end{equation}
The index $k=1,\ldots ,N$ of a site at row $r_k=1,\ldots ,N/M$ and column $c_k=1,\ldots ,M$ is given by $k=M(r_k-1) + c_k$. Conversely, we have $r_k =(\lfloor(k-1)/M\rfloor)+1$ and $c_k=\left[(k-1) \mod M\right]+1$, where $\lfloor x\rfloor$ is the floor function, i.e., $\lfloor x\rfloor$ is the largest integer not greater than x. 
The spacing between rows and columns is given by the lattice constant $a$ and therefore the interaction strength between site $k$ and $j$ reads
\begin{equation}
V_{k,j} = \frac{1}{\left(\sqrt{(r_j-r_k)^2+(c_j-c_k)^2}\right)^6}V_1.
\label{eq:mod:vkj_2d}
\end{equation}

The Hilbert space of Hamiltonian (\ref{eq:mod:gen_ham}) is spanned by the tensor product of the single-site Hilbert spaces $\mathcal{H}^{(k)}$, $\mathcal{H}= \bigotimes^N_{k=1} \mathcal{H}^{(k)}.$
The laser contributions $H_{\text{det}}$ and $H_{\text{coup}}$ are local, \ie they can be conveniently represented within $\mathcal{H}^{(k)}$. The Rydberg interaction, on the other hand, is a non-local term and connects different sites. In this case the representation is only possible in the full Hilbert space $\mathcal{H}$.
For the investigations in this work it is advantageous to define the so-called canonical product state basis,
\begin{equation}
\mathcal{S}_{\mathcal{H}}= \left\{ \ket{s_1 s_2\ldots s_N}, \quad  s_k \in \{e,g\}\right\}.
\end{equation}
In this basis the laser detuning part and the interaction part of the Hamiltonian are diagonal, whereas the laser coupling part of the Hamiltonian is off-diagonal.

For later purposes it is convenient to introduce the excitation number operator,
\begin{equation}
N_e = \sum_{k=1}^{N} n^{(k)}_e,
\end{equation}
which counts the number of Rydberg atoms in the lattice. The laser detuning part $H_{\text{det}}$ then becomes 
\begin{equation}
H_{\text{det}} = {\Delta}(N_e -N/2).
\label{eq:mod:ham_det_ne}
\end{equation}
In the basis of canonical product states the operator $N_e$ is diagonal, \ie the canonical product states $\ket{\alpha}$ are eigenstates of the operator $N_e$ with integral eigenvalues $N_e(\alpha)$. Since we consider in this work only the canonical product states as basis we remark that we usually replace the operator $N_e$ by its eigenvalue. 

\subsection{Symmetries\label{sec:mod:symmetries_ham}}
The geometry of the lattice imprints symmetries in the Hamiltonian, which we analyze in the following. The knowledge about these symmetries can be used to, \eg truncate the state space and to find degenerate states.

If the Hamiltonian $H_{\text{mod}}$ commutes with a symmetry operator $S$, the Hamiltonian conserves the symmetry, \ie it couples only states within a certain symmetry subspace. In the matrix representation of the basis of the eigenstates of the symmetry operator, this yields a block-diagonal form of $H_{\text{mod}}$.
Correspondingly, when initially preparing the system in an eigenstate of the symmetry operator, the time evolution of this state is restricted to its symmetry subspace. Hence, in such a case it is not necessary to consider the full state space. Rather it is sufficient to restrict the consideration to the corresponding subspace.

Throughout this work we presume the canonical ground state $\ket{G}=\ket{ggg\ldots g}$ as initial state. The actual relevant subspace is then determined by the symmetry properties of $\ket{G}$; depending on the number of symmetries, this reduces the number of necessary basis states and therefore the numerical effort significantly. We will give some exemplary numbers in the end of this section, but first we consider the symmetries of different lattices and indicate how to find a basis for the necessary subspace.

\paragraph{Chain}
For a linear chain with constant lattice spacing and global laser parameters we define a reflection operator $Y$ by its action on the Pauli-matrices,
\begin{equation}
Y \sigma^{(k)}_m Y^{-1} = \sigma^{(N+1-k)}_m, \quad m\in\{x,y,z\},
\end{equation}
\ie it exchanges site $k$ with site $N+1-k$. By construction we have $Y^2 = \mathbbm{1}$ and the two possible eigenvalues are $\{+1,-1\}$. Each of the 3 terms in the Hamiltonian $H_{\text{mod}}$ commutes with this operator $Y$, \ie $[H_{\text{mod}},Y]=0$.
The canonical ground state $\ket{G}$ is obviously an eigenstate to the reflection operator with eigenvalue $+1$,
${Y\ket{G}} = +1 \ket{G}$, such that it is sufficient to truncate the Hilbert space to states which are also eigenstates of $Y$ with eigenvalue $+1$, the so-called symmetric subspace.

When symmetrizing every canonical product state $\ket{\alpha}$,
\begin{equation}
 \ket{\alpha^{\text{sym}}} = \mathcal{N} \left(\mathbbm{1} + Y \right) \ket{\alpha},
\label{eq:mod:chain_symmetrized}
\end{equation}
the set of disjoint $\ket{\alpha^{\text{sym}}}$ forms a basis of the symmetric subspace.
$\mathcal{N}$ is a normalization constant such that $\braket{\alpha^{\text{sym}}}{\alpha^{\text{sym}}} = 1$ holds. The reader may easily verify that the symmetrized states are eigenstates of the reflection operator with eigenvalue $+1$. The symmetrized state $\ket{\alpha^{\text{sym}}}$ has contributions of either one or two different canonical product states depending on whether the canonical product state is already an eigenstate of the symmetry operator $Y$, \eg $ {\ket{\alpha}} = {\ket{egggee}} \rightarrow \ {\ket{\alpha^{\text{sym}}}} = \frac{1}{\sqrt{2}} \left({\ket{egggee}} + {\ket{eeggge}} \right)$ or $ {\ket{\alpha}} = {\ket{egeege}} \rightarrow \ {\ket{\alpha^{\text{sym}}}} = \ket{egeege}$.

\begin{figure}
\begin{center}
\includegraphics[width=7.5cm]{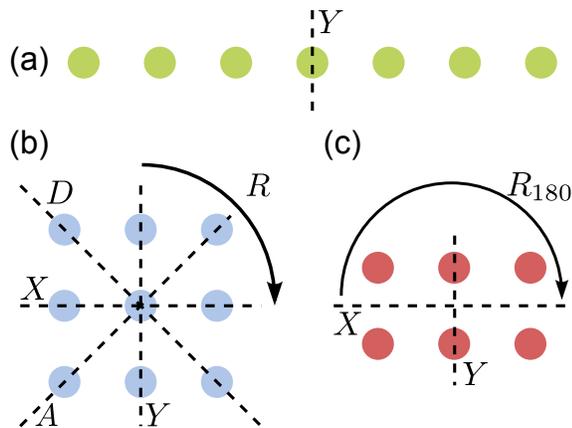}
\end{center}
\caption{(color online) Symmetry operations for (a) the chain, (b) the square, and (c) the rectangular lattice.}
\label{fig:mod:symops}
\end{figure}
\paragraph{Square}
For quadratic lattices we can define a set of symmetry operators $P=\{\mathbbm{1},R,R^2,R^3,X,D,Y,A\}$ which all commute with the Hamiltonian, $[H_{\text{mod}},S]=0,\  \forall S\in P$. These are: unity ($\mathbbm{1}$), ($p$-fold) clockwise rotations by 90 degrees ($R^p$), reflections with respect to the horizontal and vertical axis ($X,\,Y$), and reflections with respect to the diagonals ($D,\,A$). The set $P$ represents the 8 elements of the symmetry group of the square, the dihedral group $D_4$, in the state space.
In order to define the action of the operators conveniently, we momentarily replace the site index $k$ with the tuple $(r_k,c_k)$ according to its row $r_k=1, \ldots ,M$ and column $c_k=1, \ldots ,M$ in the lattice. The actions of the operators then read
\begin{align}
 R \sigma_m^{(r_k,\,c_k)} R^{-1} &= \sigma_m^{(c_k,\ M+1-r_k)}, \\
 X \sigma_m^{(r_k,\,c_k)} X^{-1} &= \sigma_m^{(M+1-r_k,\ c_k)}, \\
 D \sigma_m^{(r_k,\,c_k)} D^{-1} &= \sigma_m^{(c_k,\ r_k)}, \\
 Y \sigma_m^{(r_k,\,c_k)} Y^{-1} &= \sigma_m^{(r_k,\ M+1-c_k)},\\
 A \sigma_m^{(r_k,\,c_k)} A^{-1} &= \sigma_m^{(M+1-c_k,\ M+1-r_k)},
\label{eq:mod:sym_ops_square}
\end{align}
where $m\in\{x,y,z\}$. By definition we have $R^4=X^2=Y^2=D^2=A^2=\mathbbm{1}$. It is important to keep in mind that the mentioned operators do not commute in general. In \fig{\ref{fig:mod:symops}} we have indicated the symmetry operators. 

The canonical ground state $\ket{G}$ is an eigenstate of all 8 symmetry operators with eigenvalue $+1$,
\begin{equation}
 S \ket{g\ldots g} = +1 \ket{g\ldots g},\quad \forall \, S\in P.
\end{equation}
Accordingly, the Hilbert space can be truncated to eigenstates of all 8 symmetry operators with eigenvalue $+1$; we call this subspace symmetric as well.

By means of the group property $S P = P,\ \forall S\in P$, one finds that $S'\left(\sum_{S\in P} S \right)=\sum_{S\in P} S,\ \forall S' \in P$. Hence, for any canonical product state $\ket{\alpha}$ the superposition
\begin{equation}
\ket{\alpha^{\text{sym}}} = \mathcal{N} \left( \sum_{S\in P} S \right) \ket{\alpha}
\label{eq:mod:symmetrizing}
\end{equation}
is an eigenstate to all symmetry operators with eigenvalue $+1$, where $\mathcal{N}$ is again a normalization constant. As in the case of the chain, the resulting set of disjoint $\ket{\alpha^{\text{sym}}}$ forms a basis of the symmetric subspace.
The number of different terms in \eq{\ref{eq:mod:symmetrizing}} is 1, 2, 4 or 8 depending on whether the canonical product state $\ket{\alpha}$ is already an eigenstate to (multiple) symmetry operators in $P$. These numbers are divisors of the number of group elements in $P$ and result from group theoretical considerations in the context of orbits and stabilizers \cite{3827420180}.

\paragraph{Rectangle}
The case of the rectangular lattice can be treated analogously to the square lattice. Here, we have a set $P$ containing 4 symmetry operators $\mathbbm{1}$, $R_{180},\,X,\,Y$ which represent the symmetry group of the rectangle, the dihedral group $D_2$. $X$ and $Y$ are horizontal and vertical reflections, respectively, and $R_{180}$ is a rotation of 180 degrees. Using again the notation $(r_k,c_k)$ with $r_k=1\ldots N/M$ and $c_k=1\ldots M$, we find
\begin{align}
X \sigma_m^{(r_k,\,c_k)} X^{-1} &= \sigma_m^{(N/M+1-r_k,\ c_k)}, \\
Y \sigma_m^{(r_k,\,c_k)} Y^{-1} &= \sigma_m^{(r_k,\ M+1-c_k)},\\
R_{180} \sigma_m^{(r_k,\,c_k)} Y_{180}^{-1} &= \sigma_m^{(N/M+1-r_k,\ M+1-c_k)},
\end{align}
where $m\in\{x,y,z\}$. All four symmetry operators commute with the Hamiltonian and satisfy $R_{180}^2=X^2=Y^2=\mathbbm{1}$. The canonical ground state is again an eigenstate to all symmetry operators with eigenvalue $+1$ and \eq{\ref{eq:mod:symmetrizing}} holds with $P=\{\mathbbm{1},R_{180},X,Y\}$. Accordingly, the number of different terms in \eq{\ref{eq:mod:symmetrizing}}  for the rectangle is 1, 2 or 4.

The number of basis states of the full Hilbert space $\mathcal{H}$ is given by $2^N$, $N$ being the number of sites. As outlined before, the number of basis states reduces when considering the introduced symmetric subspaces.
In \tab{\ref{tab:num:state_space}} we provide specific examples for the number of basis states of the symmetric subspaces. For larger lattices, the state space is approximately reduced by a factor of 2, 4 and 8 for chains, rectangles and square lattices, respectively. For large square lattices, the majority of symmetrized states $\ket{\alpha^{\text{sym}}}$ consist of 8 different summands in \eq{\ref{eq:mod:symmetrizing}} since most of the canonical product states $\ket{\alpha}$ are \emph{a priori} not eigenstates to any symmetry operator. The few cases with less terms are negligible and the state space reduces by a factor of almost 8. For the chain and the rectangle the situation is similar.
\begin{table}
\begin{ruledtabular}
		\caption{Number of basis states of the full Hilbert space and of the symmetric subspace of the Hamiltonian. For the rectangular lattices we assume a square lattice, but only apply the symmetries of a rectangle to truncate the state space.\label{tab:num:state_space}}
		\begin{tabular}{lcccc}
		Number of sites & $N=4$ & $N=9$ & $N=16$ & $N=25$ \\
		\hline
		Full state space & $16$ & $512$ & $65536$ & $\sim 34\times 10^6$\\
		Linear chain & $10$ & $272$ & $32896$ & $\sim 17 \times10^6$\\
		Square lattice  & $6$ & $102$ & $8548$ & $\sim 4 \times 10^6$\\
		Rectangular lattice  & $7$ & $168$ & $16576$ & $\sim 8 \times 10^6$
 		\end{tabular}
\end{ruledtabular}
\end{table}

\section{Spectra} 
\label{sec:spectra}
\begin{figure}
\includegraphics[width=8cm]{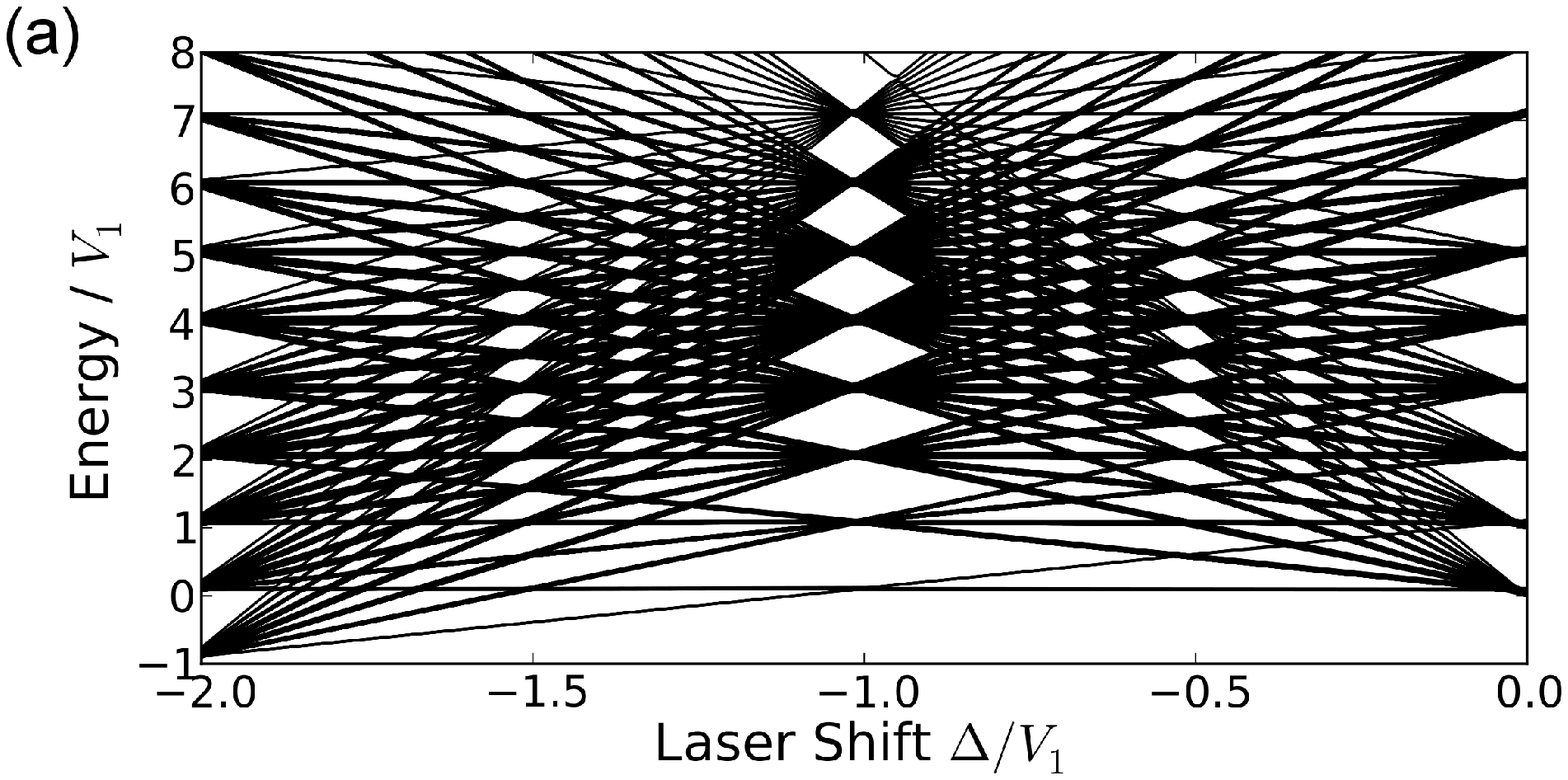}
\vspace{-0.15cm}\\
\includegraphics[width=8cm]{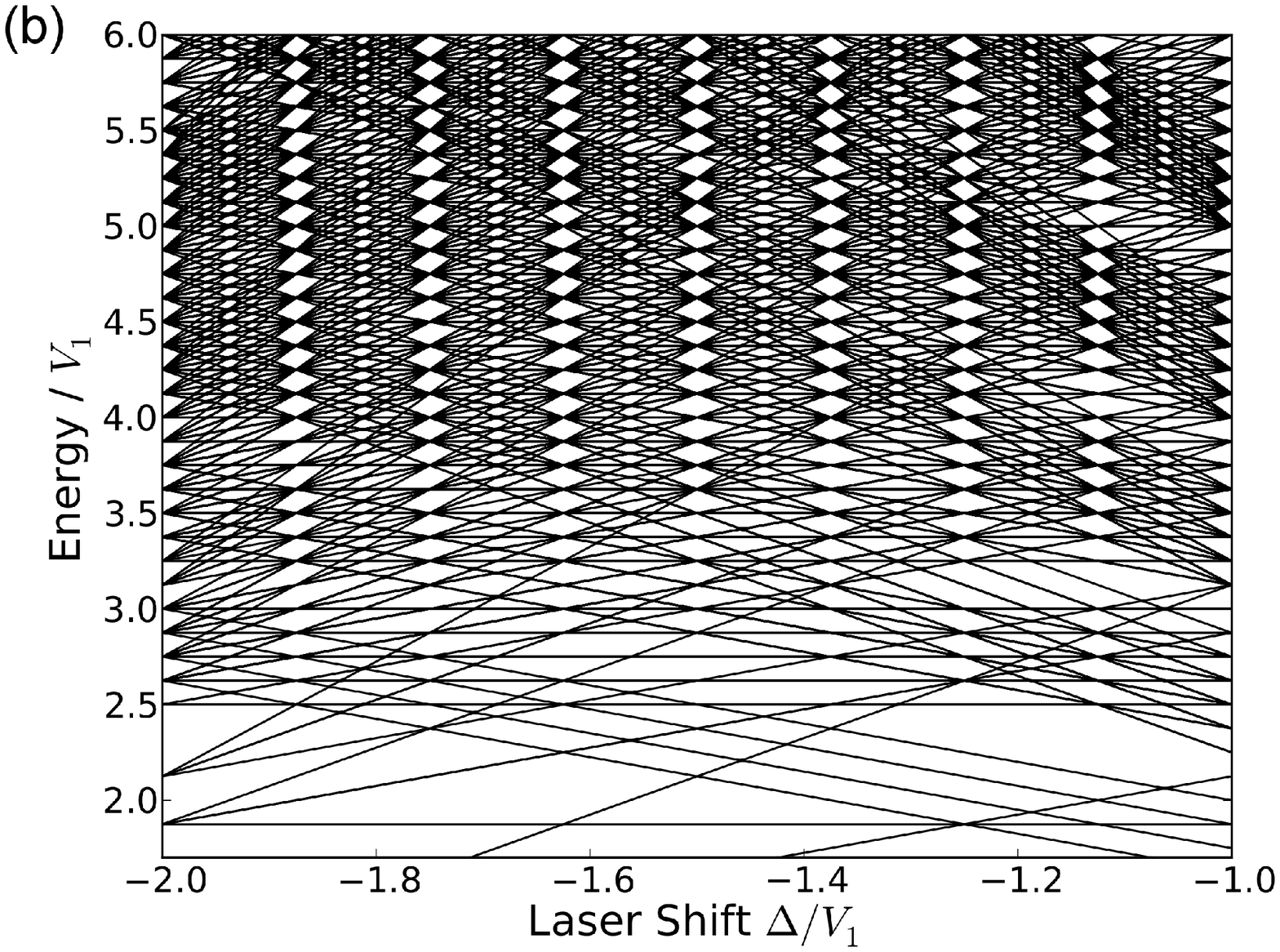}
\vspace{-0.15cm}\\
\includegraphics[width=8cm]{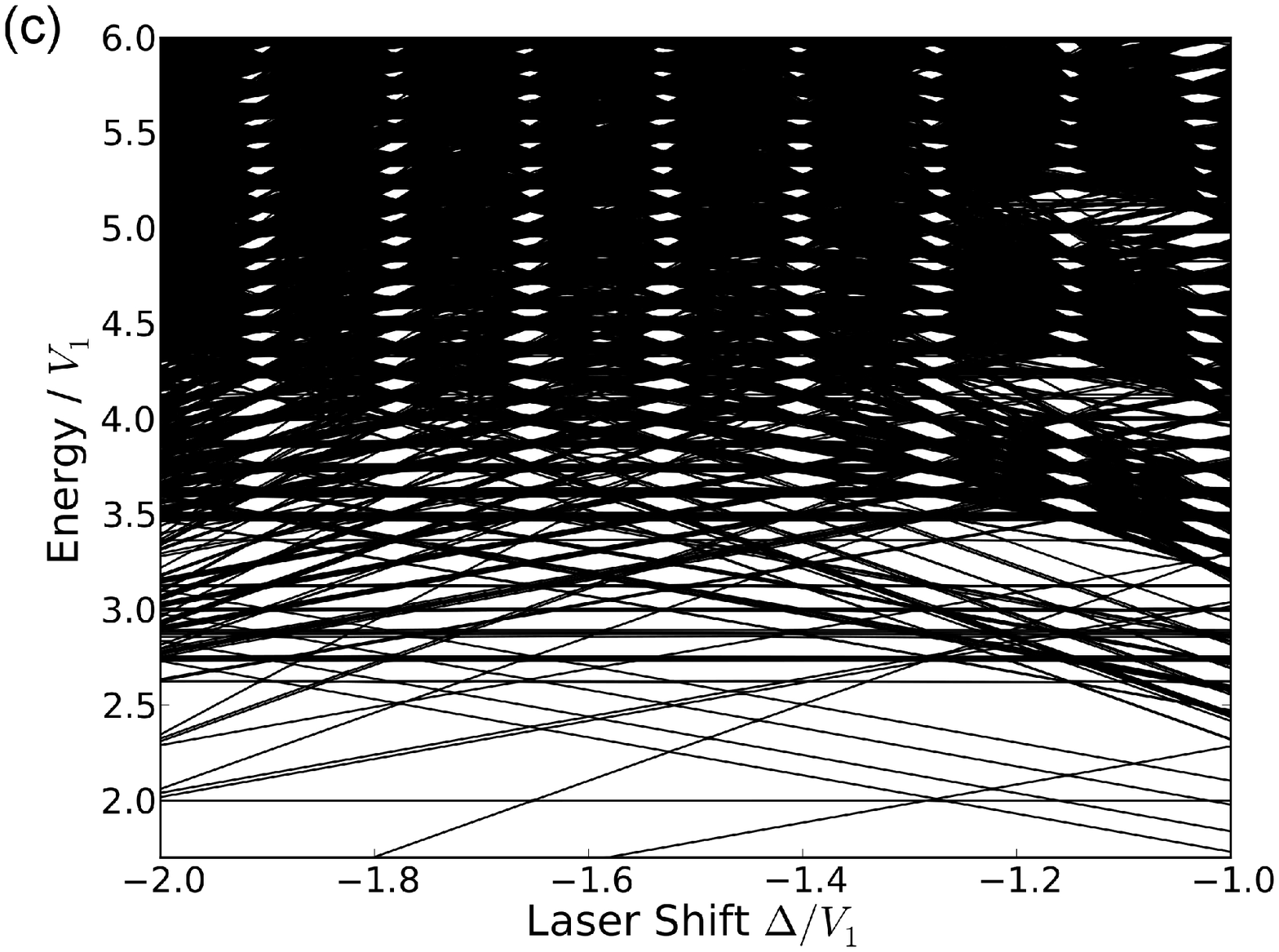}
\vspace{-0.15cm}
\caption{(a) Spectrum of the chain with 16 sites. (b,c) Spectrum of the $4\times4$ square lattice, where (b) accounts only for next- and diagonal-neighbor interactions. For the square lattice the pattern in the interaction energy occurs on a finer grid compared to the linear chain: the spectrum shows large energy gaps of $\frac18 V_1$ whenever $\Delta/V_1$ is an integral multiple of $\frac18$. Due to long-range interactions that lift the degeneracies of energy levels, this pattern is hardly visible for higher energies in the square lattice, cf.\ upper half of subfigure (c). The axes of the subfigures have different scales to ensure the visibility of the structure.
\label{fig:spec:fine_struc_square}}
\end{figure}

\setcounter{paragraph}{0}
In this section we focus on the spectra of the symmetric subspace of square lattices for a weak laser coupling, \ie $|\Omega| \ll V_1, |\Delta|$. We also compare the results with the case of a linear chain which has been already discussed thoroughly in \cite{Tezak2010a}. In the weak laser coupling regime the Hamiltonian is dominated by the diagonal contributions $H_{\text{det}}$ and $H_{\text{int}}$, while the laser coupling term $H_{\text{coup}}$ leads to a small off-diagonal perturbation. Employing \eq{\ref{eq:mod:ham_det_ne}}, the diagonal part of the Hamiltonian reads
\begin{align}
H_{\text{diag}} = {\Delta} (N_e-N/2)  + \sum_{k=1}^{N-1} \sum_{j=k+1}^{N} V_{k,j} n^{(k)}_e n^{(j)}_e
\label{eq:spec:diag_ham}
\end{align}
and its energy eigenvalue for a canonical product state $\ket{\alpha}$ is given by
\begin{equation}
\frac{E(\alpha)}{V_1} = \frac{\Delta}{V_1} (N_e(\alpha)-N/2)  + \frac{\Eint(\alpha)}{V_1},
\label{eq:spec:evals_rescaled}
\end{equation}
where $\Eint(\alpha)$ is the eigenvalue of $H_{\text{int}}$, \ie the interaction energy of the state $\ket{\alpha}$. In \fig{\ref{fig:spec:fine_struc_square}(a,c)} we show the spectrum of Hamiltonian (\ref{eq:spec:diag_ham}), \ie the $\Delta$-dependent energy eigenvalues, for a linear chain and for a square lattice. Obviously, each state appears as a straight line in the spectra since ${E(\alpha)}$ is linear in $\Delta$. Depending on the specific ratio $\Delta/V_1$ we observe points of high degeneracy (crossings) between states. According to \eq{\ref{eq:spec:evals_rescaled}}, two canonical product states $\ket{\alpha}$ and $\ket{\beta}$ with $N_e(\alpha)\neq N_e(\beta)$ become degenerate at
\begin{align}
\frac{\Delta}{V_1} &= -\frac{\Eint(\alpha)/V_1- \Eint(\beta)/V_1 }{N_e(\alpha) -N_e(\beta)}.
\label{eq:spec:slope_intmap}
\end{align}
This means that the laser detuning can be used to selectively compensate for the level shift due to Rydberg interactions in order to achieve a degeneracy between a pair of states $\ket{\alpha}$ and $\ket{\beta}$.
If $N_e(\alpha)= N_e(\beta)$ the states are either degenerate for all $\Delta$ [for $\Eint(\alpha)= \Eint(\beta)$] or never become degenerate [for $\Eint(\alpha)\neq \Eint(\beta)$].
In the following, we discuss the case of the linear chain and the square lattice in more detail.

\paragraph{Chain} The case of the linear chain has been discussed in \cite{Tezak2010a}. In the regime of weak laser coupling it is sensible to approximate the interaction energy $\Eint(\alpha)$ by next-neighbor interactions only. The number of neighboring excitations of a state $\ket{\alpha}$ is denoted by $\nee(\alpha)$, \eg $\ket{\alpha}=\ket{eeegee}\rightarrow \nee(\alpha) = 3$. According to \eq{\ref{eq:mod:vkj_chain}} neighboring excitations contribute $V_1$ to the interaction energy, thus we have $\Eint(\alpha)\approx \nee(\alpha)V_1$. \eq{\ref{eq:spec:slope_intmap}} then reads
\begin{equation}
\frac{\Delta}{V_1} = -\frac{\nee(\alpha)- \nee(\beta) }{N_e(\alpha) -N_e(\beta)}.
\end{equation}
Since the numerator and the denominator are integral numbers we observe points of degeneracy in the spectra at rational $\Delta/V_1$. The large spacing between degeneracies at integral $\Delta/V_1$ in \fig{\ref{fig:spec:fine_struc_square}(a)} is a result of the discretization of $\Eint$ in multiples of $V_1$. For fixed but integral $\Delta/V_1$ all $\alpha$-dependent terms on the right hand side of \eq{\ref{eq:spec:evals_rescaled}} are in the next-neighbor approximation integral numbers. As a consequence, also $E(\alpha)/V_1$ is discrete with spacing $1$. For rational values $\Delta/V_1=-\frac{p}{q}, \ p,q\in \mathbb{Z}\backslash\{0\}$ and $|p|,|q| \text{ coprime}$, the spacing in $E(\alpha)/V_1$ in \eq{\ref{eq:spec:evals_rescaled}} is reduced to $\frac{1}{|q|}$, which explains the smaller gaps between degeneracies visible in \fig{\ref{fig:spec:fine_struc_square}(a)} for non-integral $\Delta/V_1$. 

\paragraph{Square lattice} The situation for the square lattice is similar to that of the linear chain. However, one needs to take into account diagonal neighboring excitations to approximate the interaction energy. Analogously to $\nee(\alpha)$ we introduce the number of diagonal neighboring excitations $\neel{\sqrt{2}}(\alpha)$, which have a spatial separation of $\sqrt{2}a$. The interaction energy in next- and diagonal-neighbor approximation reads $\Eint(\alpha) = [\nee(\alpha) + \frac{1}{8} \neel{\sqrt{2}}(\alpha)]V_1$. To follow the arguments of the 1D case we introduce the effective number of diagonal neighboring excitations
\begin{equation}
N^{[\sqrt{2}]}_{ee,\text{eff}}(\alpha) := 8\nee(\alpha)  + \neel{\sqrt{2}}(\alpha),
\end{equation}
which is also an integral number. Hence, $\Eint(\alpha)=\frac{1}{8}N^{[\sqrt{2}]}_{ee,\text{eff}}(\alpha)V_1$, \ie the interaction energy is also discretized but on a finer grid with spacing $V_1/8$. Accordingly \eq{\ref{eq:spec:slope_intmap}} reads 
\begin{equation}
\frac{\Delta}{V_1} = -\frac{1}{8}\frac{N^{[\sqrt{2}]}_{ee,\text{eff}}(\alpha)- N^{[\sqrt{2}]}_{ee,\text{eff}}(\beta) }{N_e(\alpha) -N_e(\beta)},
\label{eq:spec:deg_delta_v1_square}
\end{equation}
and we expect again degeneracies between states $\ket{\alpha}$ and $\ket{\beta}$ at rational values of $\Delta/V_1$, \cf \fig{\ref{fig:spec:fine_struc_square}(b)}.
Inserting $\Delta/V_1=-\frac18 \frac{p}{q}, \ p,q\in \mathbb{Z}\backslash\{0\}$ and $|p|,|q| \text{ coprime}$ and $\Eint(\alpha)=\frac{1}{8}N^{[\sqrt{2}]}_{ee,\text{eff}}(\alpha)V_1$ in \eq{\ref{eq:spec:evals_rescaled}} we find
\begin{equation}\label{eq:Esquare}
\frac{E(\alpha)}{V_1} = \frac18\left[-\frac{p}{q} (N_e(\alpha)-N/2) + N^{[\sqrt{2}]}_{ee,\text{eff}}(\alpha)\right].
\end{equation}
Since $N_e(\alpha)$ and $N^{[\sqrt{2}]}_{ee,\text{eff}}(\alpha)$ are integers we expect a spacing in ${E(\alpha)}/{V_1}$ of $\frac18 \frac{1}{|q|}$. Equation (\ref{eq:Esquare}) describes why we observe much smaller energy gaps between degeneracies in square lattices and that energy gaps of $\frac18V_1$ appear at multiples of $1/8$ for $\Delta/V_1$, \cf \fig{\ref{fig:spec:fine_struc_square}(b)}.

\begin{figure}
\begin{center}
\includegraphics[width=8.5cm]{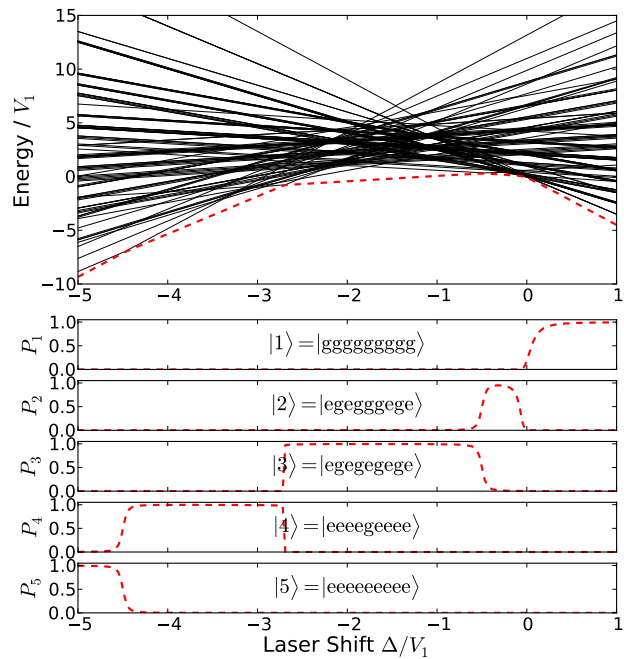} 
\end{center}
\caption{(color online) Spectrum of the symmetric subspace of Hamiltonian (\ref{eq:mod:gen_ham}) for the square lattice with $3 \times 3$ sites with a laser coupling of $\Omega = 0.05V_1$. The lower half of the plot shows the projection probabilities of the ground state (red, dashed) on a sample of (symmetrized) basis states.}
\label{fig:spec:N_9_M_3_gs}
\end{figure}

In general, $N_e$, $\nee$, and $\neel{\sqrt{2}}$ are restricted by the geometry of the lattice and not all combinations are possible. As a result we partially observe larger gaps between degeneracies in the lower half of \fig{\ref{fig:spec:fine_struc_square}(b)} than expected from our previous argumentation. Specifically, due to the small size of the $3\times3$ lattice, its spectrum shows on the one hand the fine structure of the square lattice and on the other hand still large gaps at integral $\Delta/V_1$ like in the case of the chain, \cf \fig{\ref{fig:spec:N_9_M_3_gs}}. 

Let us now have a look at the influence of the long-range interactions neglected so far. In \fig{\ref{fig:spec:fine_struc_square}}(b,c) we compare the spectra of the $4\times 4$ lattice for next- and diagonal neighbor interactions to the one including the long-range interactions. Although still visible, the structure in the spectrum is less developed, \ie the gaps diminish, especially in the upper half of subfigure (c). This can be explained as follows: The contributions of long-range interactions wash out the discretization of the interaction energy $\Eint(\alpha)=\frac{1}{8}N^{[\sqrt{2}]}_{ee,\text{eff}}(\alpha)V_1$. Therefore the lines in the spectrum (b) turn partially into a large number of narrow lines which are recognizable as thick lines in subfigure (c). Points of high degeneracy turn into regions with a high density of states and a large number of degeneracies. As a result also the spacing in $E(\alpha)$ is washed out and the gap sizes in the spectrum diminish. As we can see in the upper half of subfigure \fig{\ref{fig:spec:fine_struc_square}}(c), already for the $4\times4$ lattice the deviations due to long-range interactions are large enough to partly fill even the gaps of $\frac18 V_1$ in the spectrum. We expect that these effects become stronger with increasing lattice size due to an increasing number of long-range interactions.
The same applies in principle also to the linear chain. However, in this case the effects of long-range interactions are smaller compared to square lattices since the gaps in the chain are larger (up to $V_1$) while the next-to-next-neighbor contributions are of order $V_1/64$. In square lattices we usually have gaps of $V_1/8$ but the next-to-diagonal-neighbor contributions are of order $V_1/64$, as well. In addition, these contributions are more numerous in the case of the square lattice.
We remark that the long-range interactions shift windows in the spectrum slightly to more negative values of $\Delta/V_1$, \cf \fig{\ref{fig:spec:fine_struc_square}(c)}.

\section{Ground states}
\label{sec:ground_states}

In this section we focus on the laser detuning-dependent ground states of the symmetric subspace of the system. This is of interest, e.g., for an adiabatic sweep through the ground states to selectively prepare a desired state \cite{PhysRevLett.104.043002}.
In \fig{\ref{fig:spec:N_9_M_3_gs}} we indicate the energetic ground state in the spectrum of the $3\times3$ lattice with a dashed, red line. The associated projection probabilities $P_i$, also shown in \fig{\ref{fig:spec:N_9_M_3_gs}}, have been obtained by exact diagonalization and obviously depend on the laser detuning $\Delta/V_1$. In general, a numerical diagonalization is not feasible due to the exponential growth of the state space. We therefore consider the limit of a vanishing laser coupling, $\Omega=0$, which avoids the diagonalization since the resulting Hamiltonian
\begin{equation}
H_{\text{diag}} = \frac{\Delta}{2} \sum_{k=1}^{N}\sigma^{(k)}_z + \sum_{k=1}^{N-1} \sum_{j=k+1}^{N} V_{k,j} n^{(k)}_e n^{(j)}_e
\label{eq:gs:diag_ham}
\end{equation}
is diagonal in the basis of canonical product states. Using Hamiltonian (\ref{eq:gs:diag_ham}), the projection probabilities in \fig{\ref{fig:spec:N_9_M_3_gs}} turn into step functions rather than having smooth edges. The limit of vanishing laser coupling is expected to be a reasonable approximation for ground states in the presence of the driving laser whenever the ground state is energetically well separated from other states it can possibly interact with. This is often the case for what we will call dominant ground states in Sec.\ \ref{sec:gs:seq} and which are marked by bold numbers in Tables \ref{tab:seq:chain15}-\ref{tab:seq:rectangle24-8}. There are, however, also dominant ground states states that are nearly degenerate with excited states [see \fig{\ref{fig:seq:gs_chains}}(d) for an example] and hence are more sensitive to a non-vanishing laser coupling. Similarly, excited states come close in energy in the crossover regime between dominant ground states. There, one finds a superposition of canonical states, resulting in the smooth, non-unitary projection probabilities found in \fig{\ref{fig:spec:N_9_M_3_gs}}. A more quantitative study of the effect of a non-vanishing laser coupling is beyond the scope of the present work and is also specific to the particular lattice geometry and dimension considered.

Even in the case of vanishing laser coupling the determination of the ground states is non-trivial, if one refrains from evaluating the complete diagonal, which also becomes inefficient with increasing lattice size. A more sophisticated method to obtain the ground states of \eq{\ref{eq:gs:diag_ham}} is to solve the problem via \emph{maximum cuts} as we will explain in the next section. Prior to that we summarize some general properties of the ground states. As in the previous section, the eigenvalues of $H_{\text{diag}}$ are given by \eq{\ref{eq:spec:evals_rescaled}}. When searching for the ground state for fixed $\Delta/V_1$ we can neglect the constant offset $-N\Delta/(2V_1)$. Hence, we are interested in the canonical product state $\ket{\alpha}$ which minimizes the energy
\begin{equation}
\frac{E(\alpha)}{V_1} = \frac{\Delta}{V_1}N_e(\alpha)  + \frac{\Eint(\alpha)}{V_1}.
\label{eq:gs:energy}
\end{equation}

To achieve this, we divide the canonical product states $\ket{\alpha}$ into groups according to their number of excitations $N_e(\alpha)$ such that within each group $N_e$ is fixed. Thus, for each group the term $N_e(\alpha){\Delta}/{V_1}$ in \eq{\ref{eq:gs:energy}} is constant and minimizing $E(\alpha)$ for fixed laser detuning $\Delta$ implies minimizing the interaction energy $\Eint(\alpha)$. As a result, the configuration of minimum interaction energy (MEC) for each possible $N_e=0,\ldots,N$ is a ground state candidate; vice versa, every ground state is a MEC. Determining the true ground state from the $N+1$ MECs is then a trivial task. Finding the $N+1$ MECs, however, is a non-trivial optimization problem. We remark that there is in general more than one MEC per $N_e$ since states are usually degenerate for symmetry reasons or might be even accidentally degenerate. From \eq{\ref{eq:gs:energy}} it is evident that for $\Delta/V_1>0$ the ground state is given by the canonical ground state $\ket{G}$ since $\Eint(\alpha)\geq 0$. With decreasing $\Delta/V_1$ the number of excitations of the ground state increases monotonically and for large negative detunings, $\Delta/V_1\ll 0$, the fully excited lattice $\ket{ee\ldots e}$ becomes the ground state. See Fig.~\ref{fig:spec:N_9_M_3_gs} for an example.

\subsection{Ground states and max-cut}
\label{sec:gs:max-cut}

In this subsection we show that the evaluation of the ground state of \eq{\ref{eq:gs:diag_ham}} for a certain $\Delta/V_1$  is equivalent to a combinatorial optimization problem, the \emph{maximum cut} problem (max-cut). A sophisticated code for solving max-cut allows then to obtain the ground states for a large number of systems. For the formulation as max-cut problem it is necessary to transfer our Hamiltonian to the Ising-model form,
\begin{align}
H_{\text{Is}}(\omega) &=  -\sum^N_{k=1} h_k S_k - \sum^{N-1}_{k=1}\sum^N_{j=k+1} J_{k,j} S_k S_j.
\label{eq:gs:ising_model}
\end{align}
Here, the spins take on only two discrete values $S_k=\pm1$ and the spin configuration is indicated by $\omega$. The strength of the interaction is given by real weights $J_{k,j}$ and $h_k$ is a site-dependent external magnetic field.

For the transformation to the Ising model form, we first rewrite $H_{\text{diag}}$ in terms of Pauli matrices. In appendix \ref{sec:app:appendix} we demonstrate that using $n^{(k)}_e = [\sigma^{(k)}_z+ \mathbbm{1}]/2$ and rearranging the terms of the interaction part of the Hamiltonian yields
\begin{align}
\nonumber H_{\text{diag}} =&  \sum_{k=1}^{N} \left( \frac{\Delta}{2} + \frac14 \sum_{j\neq k} V_{k,j} \right)  \sigma^{(k)}_z\\ & + \sum_{k=1}^{N-1} \sum_{j=k+1}^{N} \frac14  V_{k,j} \sigma^{(k)}_z \sigma^{(j)}_z+\mathrm{const.},
\label{eq:gs:diag_sigmaz}
\end{align}
where $V_{k,j}=V(|\mathbf{r}_k-\mathbf{r}_j|)$. The desired Ising model form (\ref{eq:gs:ising_model}) can be obtained by translating $\sigma^{(k)}_z \rightarrow S_k$ while omitting the constant offset:
\begin{align}
H_{\text{Is}} &=  \sum^N_{k=1} \left( \frac{\Delta}{2} + \frac14 \sum_{j\neq k} V_{k,j} \right) S_k + \sum^{N-1}_{k=1}\sum^N_{j=k+1} \frac14 V_{k,j} S_k S_j.
\label{eq:gs:diag_ising_form}
\end{align}
In a physical interpretation, the Rydberg-Rydberg interaction leads to an additional site-dependent external magnetic field in the Ising model.

If we define a fixed `ghost spin' $S_0=+1$ and put $J_{0,j}=h_j$, Hamiltonian (\ref{eq:gs:diag_ising_form}) can be written in the compact form
\begin{align}
H_{\text{Is}}(\omega) = -\sum^{N-1}_{k=0}\sum^N_{j=k+1} J_{k,j} S_k S_j,
\label{eq:gs:ising_with_ghost}
\end{align}
with
\begin{align}
\label{eq:gs:j_ising_0} J_{0,j} &= -\left(\frac{\Delta}{2} + \frac14 \sum^N_{\substack{i=1\\i\neq j}} V_{j,i}\right), \qquad j=1,\ldots, N,\\
\label{eq:gs:j_ising_k} J_{k,j} &= -\frac14 V_{k,j}, \qquad k\neq j,\ k,j =1,\ldots, N.
\end{align}

As presented in \cite{9783527404063} the ground states of such an Ising Hamiltonian can be obtained by means of the concept of max-cuts in graph theory. For this, the individual spins $k$ in the lattice are represented by vertices (or nodes) $k$ in the vertex set $V$ of a graph $G$. The interaction of two spins $k,j$ is represented by the edge $(k,j)$ connecting the nodes $k$ and $j$, its ends, in the graph. The interaction strength is taken into account by a real edge weight $c_{k,j}=-J_{k,j}$. The choice $S_k\in\{+1,-1\}$ divides the set of nodes $V$ into two (disjoint) subsets $V^+=\{k\in V | S_k=+1\}$ and $V^-=\{k\in V | S_k=-1\}$, respectively. A cut $\delta(V^+)=\delta(V^-)$ is given by the set of edges, where one end is in $V^+$ and the other in $V^-$; the weight of a cut is given by the sum of the weights of all its edges. As demonstrated in \cite{9783527404063}, minimizing \eq{\ref{eq:gs:ising_with_ghost}} is equivalent to finding a cut with maximum weight.
We add some remarks:
\begin{itemize}
 \item The idea given above is valid for a large class of systems since the geometry and the interaction potential enter only via the interaction potential $V_{k,j}=V(|\mathbf{r}_k-\mathbf{r}_j|)$.
 \item Max-cut will only return one configuration $\omega$ for which $H_{\text{Is}}(\omega)$ reaches a minimum. However, in general this configuration is degenerate with others and it is not predictable which of them max-cut will return. If the degeneracies occur for symmetry reasons, obtaining the other configurations is possible by applying the symmetry operators of the lattice, \cf \sect{\ref{sec:mod:symmetries_ham}}. An additional accidental degeneracy (not for symmetry reasons) remains undetected in general. 
 \item The max-cut problem is in general NP-hard \cite{9783527404063} and therefore also challenging. 
\end{itemize}

Using max-cut to determine all ground states of the system requires to solve the max-cut problem many times in a complete parameter range of $\Delta/V_1$. We scanned the parameter range recursively and the recursion was stopped by limiting the maximal resolution in $\Delta/V_1$ to a certain threshold, \eg $10^{-6}$. This means if a MEC becomes a ground state for a range in $\Delta/V_1$ which is below this threshold, we will in general not detect this MEC. The same applies to configurations which never become ground state of the system. The latter case appears for example in 2D lattices. Nevertheless, for most of the below considered systems we can present the complete set of $N+1$ MECs, because we additionally determined the MECs with a brute force method. For this the complete Hamiltonian diagonal is evaluated which takes for 36 sites on a single-core workstation about 2 weeks and the computation time roughly doubles with each additional site. In comparison, a corresponding max-cut instance can be solved in only 10 minutes, being an impressive enhancement such that max-cut allowed us to determine the MECs of lattices with up to 49 sites.

\subsection{Ground state sequences}
\label{sec:gs:seq}

In the previous section we introduced our method to search for ground states. In this section, we present results for selected examples. We start with a brief discussion for the linear chain and then turn to the ground states of the square lattices. Subsequently we summarize some properties we found in the case of a rectangular lattice and show two examples. 

In the following, we will often talk about MECs: as a function of the laser detuning $\Delta$, the ground state is a sequence of MECs, each of which is the ground state of the system for a given range in $\Delta$.

\subsubsection{Chain\label{sec:gs:chain}}
We start with the discussion of the ground states of the linear chain. For the linear chain we are able to calculate the MECs for up to 36 sites with max-cut and verified them by the brute force method (double precision: $\sim 10^{-15}$). The difficulties for the max-cut code to find the ground state of larger linear chains stem from the very small interaction contribution if the excitations are far apart from each other, \eg the smallest one being $V_1/(N-1)^6$. The necessary accuracy to distinguish between a ground state and an excited state poses a numerical challenge and therefore reduces the possible/feasible chain sizes.

\begin{figure}
\begin{center}
\includegraphics[width=8cm]{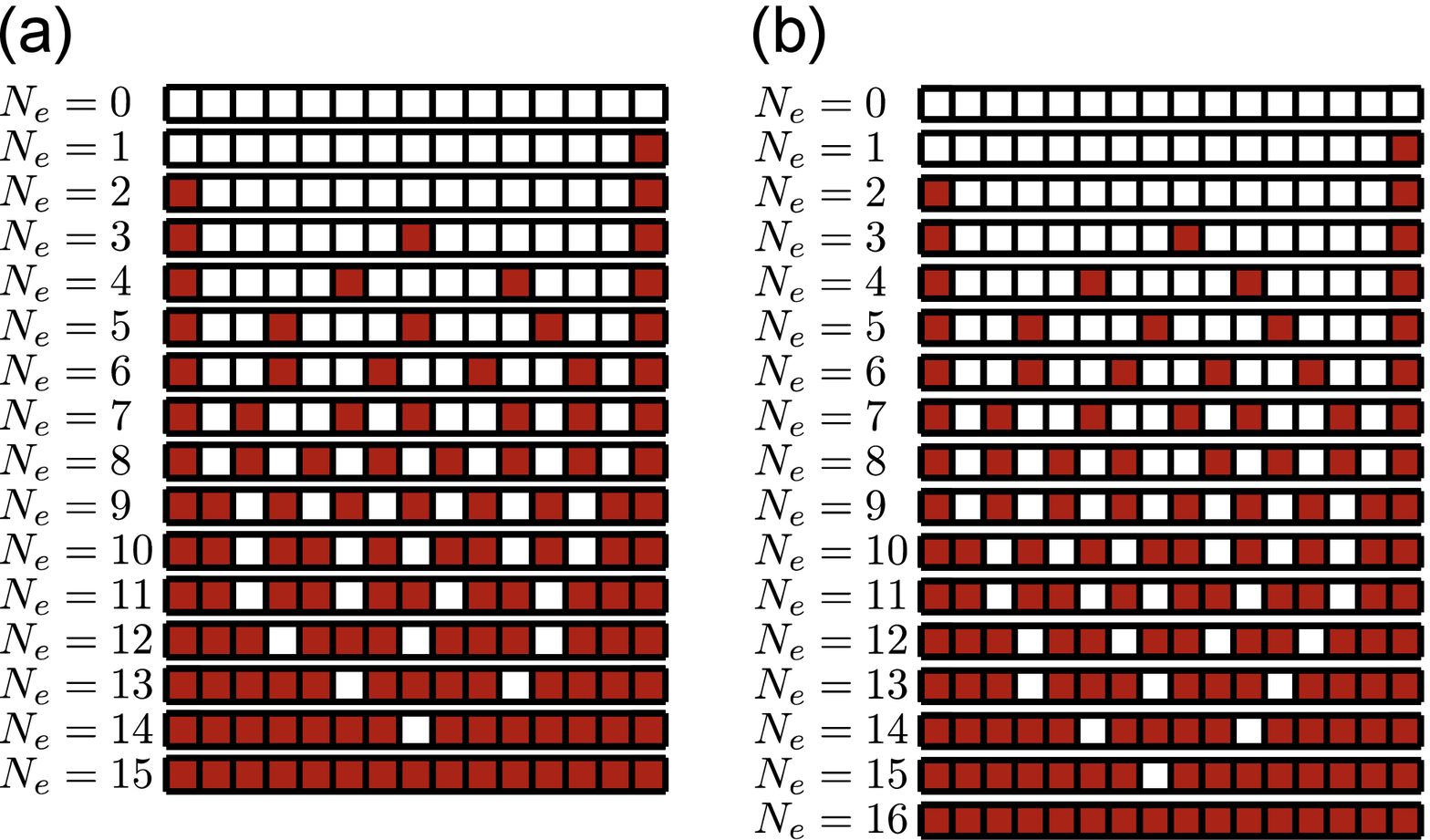}
\includegraphics[width=8cm]{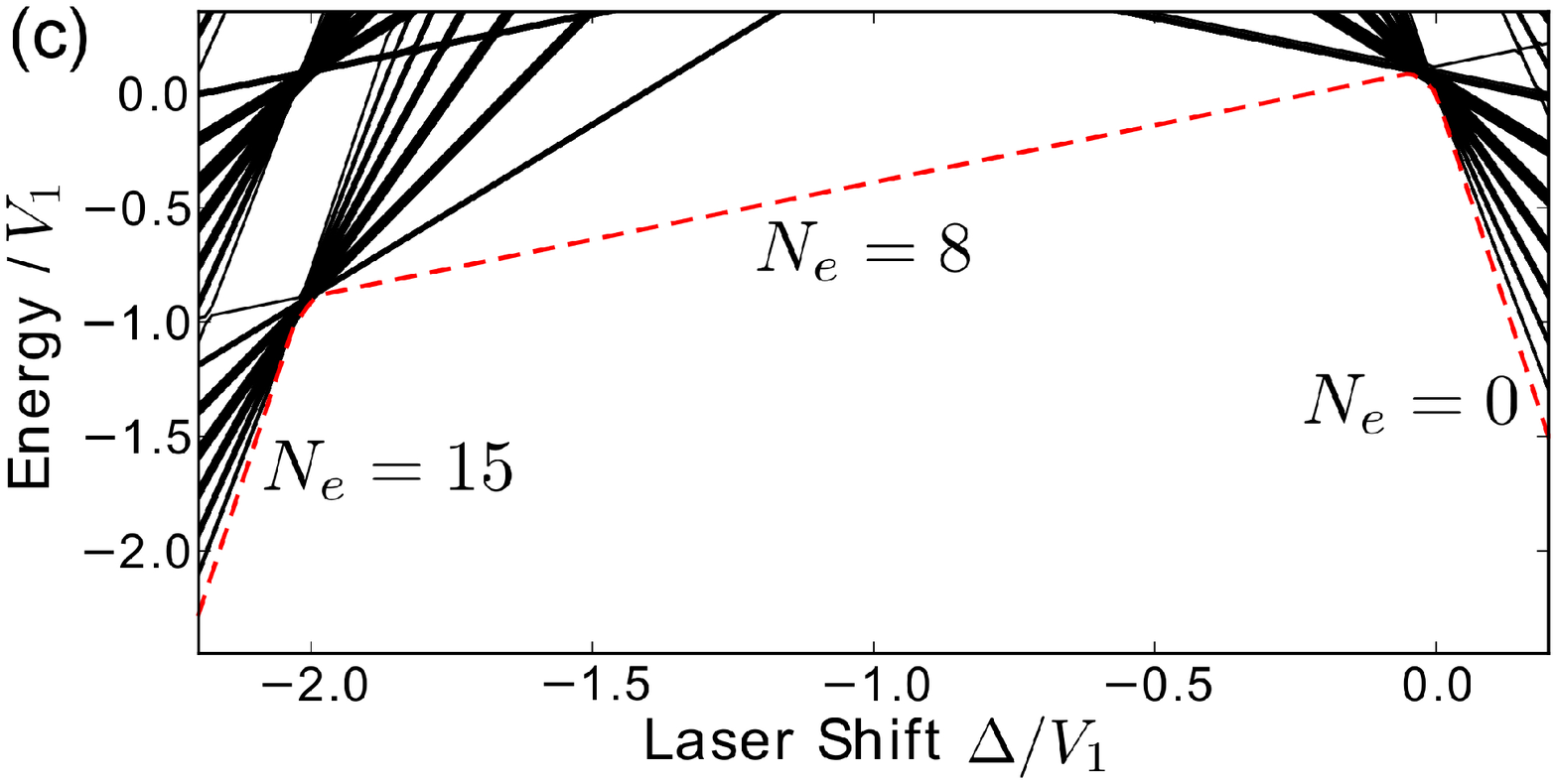}
\includegraphics[width=8cm]{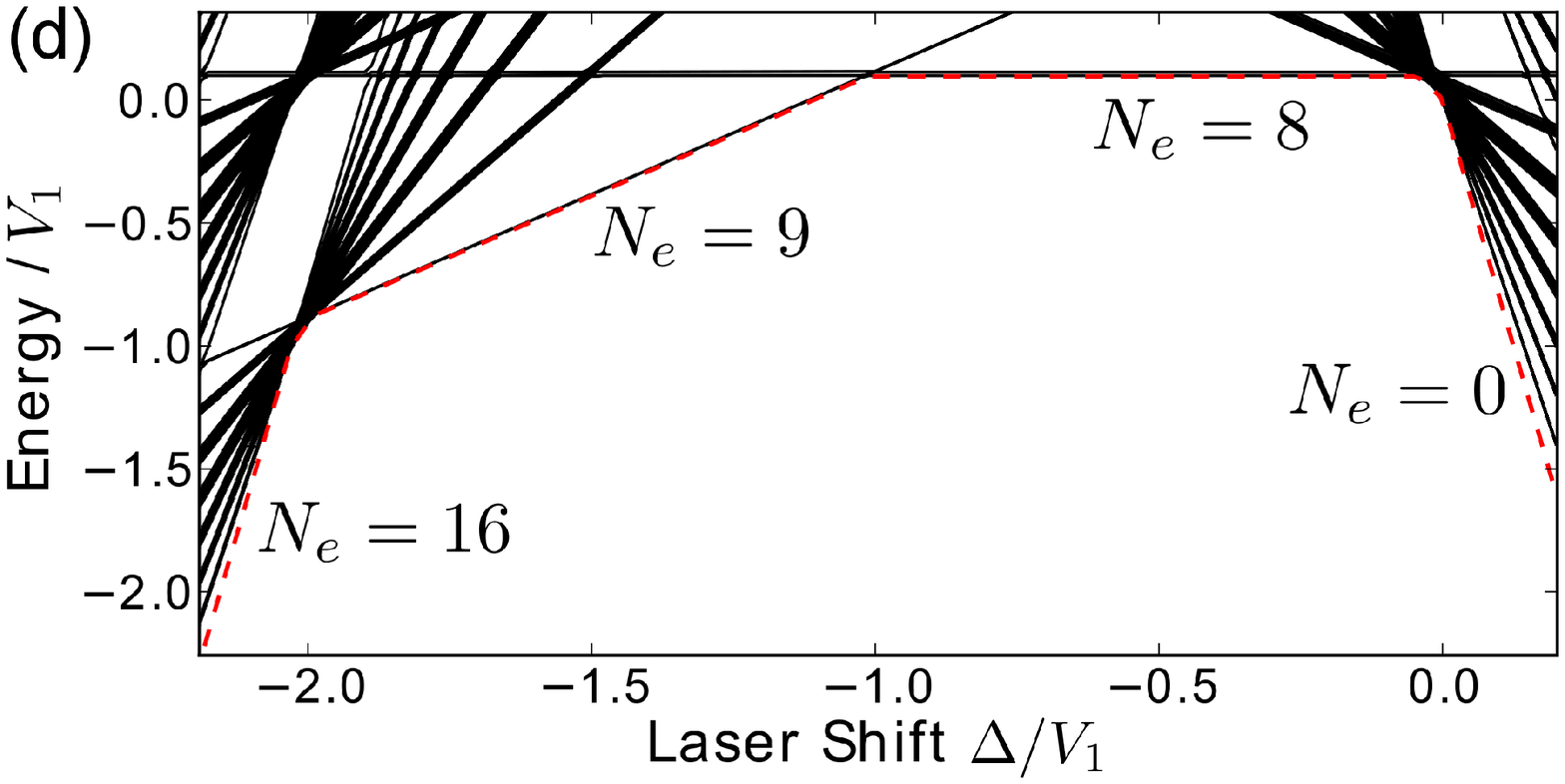}
\end{center}
\caption{(color online) Illustration of the MECs/ground states of the linear chains with $N=15$ (a) and $N=16$ (b). Excitations are filled (red/gray), ground state atoms in white. Note, these states are usually degenerate for symmetry reasons. (c) and (d) show a detail of the spectrum, where the ground state is indicated by a red, dashed line.}
\label{fig:seq:gs_chains}
\end{figure}

In \fig{\ref{fig:seq:gs_chains}(a,b)} we present the ground state MECs for the chains with $N=15$ and $N=16$. The crossover laser detunings where the ground state switches from one MEC to another are given in Tables \ref{tab:seq:chain15} and \ref{tab:seq:chain16}. Note that these states are usually degenerate with other canonical product states due to the reflection symmetry of the chain; the ground states of the \emph{symmetric} subspace can be obtained by employing \eq{\ref{eq:mod:chain_symmetrized}}. Product states with $N_e=1$ are exceptions because a single excitation can be placed on an arbitrary site without changing the interaction energy. Thus, the $N$ canonical product states with $N_e=1$ are all degenerate and the ground state of the symmetric subspace is a superposition of symmetrized product states with $N_e=1$. Besides the symmetry-induced degeneracies, in general it might be possible to have states being accidentally degenerate with the obtained MECs. 

\begin{table}
\begin{ruledtabular}
	\begin{center}
		\caption{Chain with $N=15$: Range for which the MECs in \fig{\ref{fig:seq:gs_chains}(a)} become a ground state. For example, the MEC with $N_e=15 $ becomes a ground state in the range $-\infty < \Delta/V_1\leq -2.034670$.\label{tab:seq:chain15}}
		\begin{tabular}{cc|cc|cc}
		$\Delta/V_1$ & $N_e$ & $\Delta/V_1$ & $N_e$ & $\Delta/V_1$ & $N_e$ \\
		\hline
		$-\infty$ & \textbf{15} & -1.989622 & 9 & -0.000358 & 3 \\
		-2.034670 & 14 & -1.986998 & \textbf{8} & -0.000017 & 2 \\ 
		-2.034494 & 13 & -0.045188 & 7 & 0.000000 & 1 \\
		-2.033691 & 12 & -0.044529 & 6 & 0 & \textbf{0} \\
		-2.029601 & 11 & -0.017995 & 5 & $\infty$ &   \\
		-2.003177 & 10 & -0.002879 & 4 & & \\
		-1.989622 &    & -0.000358 &   & &
		\end{tabular}
	\end{center}
\end{ruledtabular}
\end{table}
\begin{table}
\begin{ruledtabular}
	\begin{center}
		\caption{Chain with $N=16$: Range for which the MECs in \fig{\ref{fig:seq:gs_chains}(b)} become a ground state.\label{tab:seq:chain16}}
		\begin{tabular}{cc|cc|cc}
		$\Delta/V_1$ & $N_e$ & $\Delta/V_1$ & $N_e$ & $\Delta/V_1$ & $N_e$ \\
		\hline
		$-\infty$ & \textbf{16} & -1.989913 & 10 & -0.001927 & 4 \\
		-2.034673 & 15 & -1.988232 & \textbf{9} & -0.000182 & 3 \\ 
		-2.034555 & 14 & -1.016109 & \textbf{8} & -0.000012 & 2 \\ 
		-2.033875 & 13 & -0.045014 & 7 & 0.000000 & 1 \\ 
		-2.030749 & 12 & -0.044340 & 6 & 0 & \textbf{0} \\ 
		-2.016152 & 11 & -0.004830 & 5 & $\infty$ &  \\ 
		-1.989913 &    & -0.001927 &   & &
		\end{tabular}
	\end{center}
\end{ruledtabular}
\end{table}

The MECs in \fig{\ref{fig:seq:gs_chains}(a,b)} seem to trivially separate the excitations as far as possible, especially for $N_e<N/2$. However, due to the interplay of many long-range interactions, it is hard to predict the configuration of minimum interaction energy, in particular if the chain size increases. Hence, we will not further discuss the excitation patterns of the MECs but consider them in the following as given.

Empirically, we find for all investigated linear chains that for each number of Rydberg atoms $N_e$ there is a MEC which becomes the ground state for some laser detuning $\Delta/V_1$. This is in general not the case for other lattices as we will show later. However, most of the ground states appear only for very small intervals in $\Delta/V_1$ and could be therefore hard to detect in an experimental realization. In other words, while varying the laser detuning we find regions where the ground state changes rapidly (clustered ground state crossovers) and regions where the ground state does not change in a wide range, \cf dashed line in \fig{\ref{fig:seq:gs_chains}(c,d)}. For the examples $N=15$ and $N=16$, most dominant ground states are the MECs with $N_e=8$ and $N_e=8,\,9$, respectively and trivially $N_e=0$ and $N_e=N$. This is also visible in \fig{\ref{fig:seq:gs_chains}(c,d)} and in Tables \ref{tab:seq:chain15} and \ref{tab:seq:chain16} (bold numbers). 

In general we observed that for the chain the MECs with $N_e=\lceil \frac{N}{2} \rceil$ and $N_e=\lceil \frac{N+1}{2} \rceil$ become ground states in a wide range in $\Delta/V_1$, where $\lceil x\rceil$ is the ceiling function, i.e., $\lceil x\rceil$ is the smallest integer not less than x. To qualitatively understand this, we have a closer look at the energy difference between two subsequent MECs $\ket{\alpha},\, \ket{\beta}$ with $N_e(\alpha)$ and $N_e(\beta)=N_e(\alpha)+1$ excitations, respectively. The laser detuning where $E(\beta) = E(\alpha)$, \ie for which the ground state changes from the MEC with $N_e(\alpha)$ to $N_e(\beta)$, reads according to \eq{\ref{eq:spec:slope_intmap}}
\begin{align}
 {\Delta}_{N_e(\alpha),N_e(\alpha)+1} = - \left[\Eint(\beta) - \Eint(\alpha)\right],
\label{eq:seq:energy_incr_next}
\end{align}
where the ground state for $\Delta< \Delta_{N_e(\alpha),N_e(\alpha)+1}$ is represented by $\ket{\beta}$ and for $\Delta > \Delta_{N_e(\alpha),N_e(\alpha)+1}$ by $\ket{\alpha}$.

For the case of $N=15$ we consider the MECs with $N_e=7,8,9,10$. The associated interaction energies are $\Eint/V_1\approx 0,0,2,4$, respectively, approximately given by the number of neighboring excitations. Hence, we arrive at the crossover detunings $\Delta_{7,8}/V_1\approx0$, $\Delta_{8,9}/V_1\approx-2$ and $\Delta_{9,10}/V_1\approx-2$ agreeing with \tab{\ref{tab:seq:chain15}} and the wide range in \fig{\ref{fig:seq:gs_chains}(c)} for which the MEC with $N_e=8$ is a ground state. For $N=16$ we find $\Eint/V_1\approx 0,0,1,3$ for the MECs with $N_e=7,8,9,10$. This yields $\Delta_{7,8}/V_1\approx0$, $\Delta_{8,9}/V_1\approx -1$ and $\Delta_{9,10}/V_1\approx -2$, \cf \tab{\ref{tab:seq:chain16}} and \fig{\ref{fig:seq:gs_chains}(d)}. 

In general, if we assume that the excitations of the MEC with $N_e=\frac{N+1}{2}$ of a chain with an odd number of sites $N$ are equidistantly distributed, this excitation pattern has $\nee=0$, \ie $\Eint/V_1\approx 0$. Therefore the following MEC has $\nee=2$, $\Eint/V_1\approx 2$ and the alternating chain is a dominant ground state. Similar arguments are possible for the case of an even number of sites.

\subsubsection{Square lattice}

\begin{figure}
\begin{center}
\includegraphics[width=6cm]{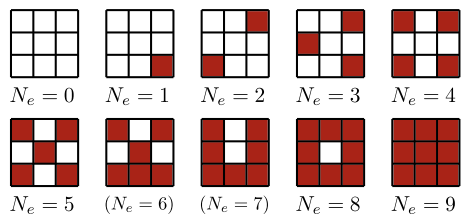}
\end{center}
\caption{(color online) Illustration of the MECs/ground states of the square with $N=9$. Excitations are filled (red/gray), ground state atoms in white. Note, these states are usually degenerate for symmetry reasons. Parentheses indicate MECs which do not appear in the ground state sequence. These have not been found by max-cut but by a brute force method.}
\label{fig:seq:square9}
\end{figure}
\begin{table}
\begin{ruledtabular}
	\begin{center}
		\caption{Square with $N=9$: Range for which the MECs in \fig{\ref{fig:seq:square9}} become a ground state. Bold numbers indicate states which are ground state for a wider range in $\Delta$ (`dominant ground states') and the lack of a certain $N_e$ in the ground state sequence is emphasized by underlining the next higher $N_e$.\label{tab:seq:square9}}
		\begin{tabular}{cc|cc|cc}
		$\Delta/V_1$ & $N_e$ & $\Delta/V_1$ & $N_e$ & $\Delta/V_1$ & $N_e$ \\
		\hline
		$-\infty$ & \textbf{9} & -0.5      & 4 & -0.001953 & 1 \\
		-4.5      & \underline{\textbf{8}} & -0.034781 & 3 & 0 & \textbf{0} \\
		-2.698417 & \textbf{5} & -0.029672 & 2 & $\infty$ & \\
		-0.5      &   & -0.001953 & & &
		\end{tabular}
	\end{center}
\end{ruledtabular}
\end{table}

\begin{figure}
\begin{center}
\includegraphics[width=8cm]{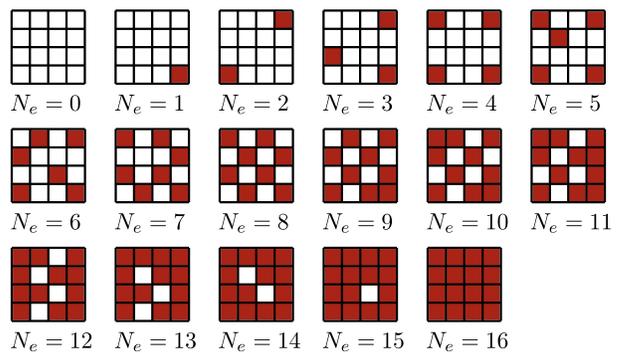}
\end{center}
\caption{(color online) Same as in \fig{\ref{fig:seq:square9}} but for $N=16$.}
\label{fig:seq:square16}
\end{figure}

\begin{figure}
\begin{center}
\includegraphics[width=8cm]{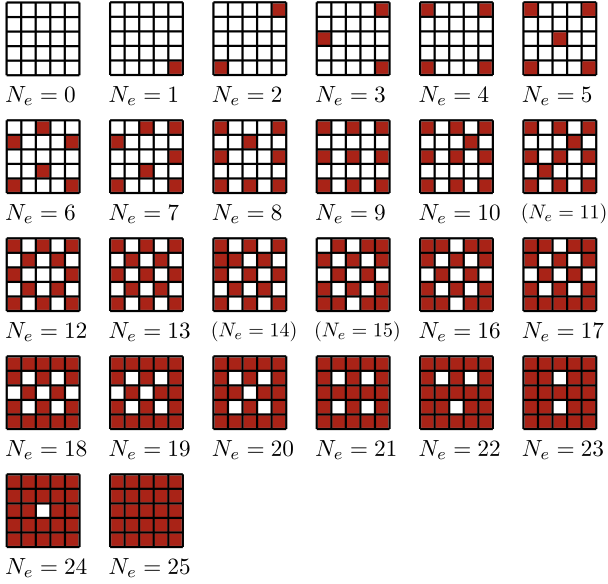}
\end{center}
\caption{(color online) Same as in \fig{\ref{fig:seq:square9}} but for $N=25$.}
\label{fig:seq:square25}
\end{figure}

\begin{figure}
\begin{center}
\includegraphics[width=8cm]{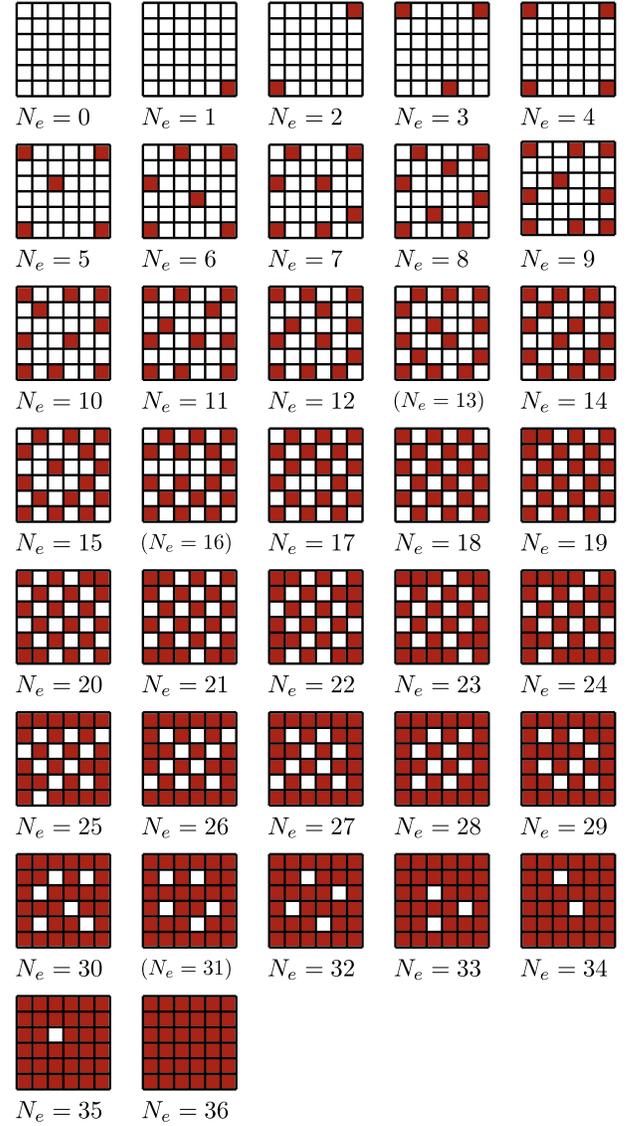}
\end{center}
\caption{(color online) Same as in \fig{\ref{fig:seq:square9}} but for $N=36$.}
\label{fig:seq:square36}
\end{figure}

\begin{figure}
\begin{center}
\includegraphics[width=8.5cm]{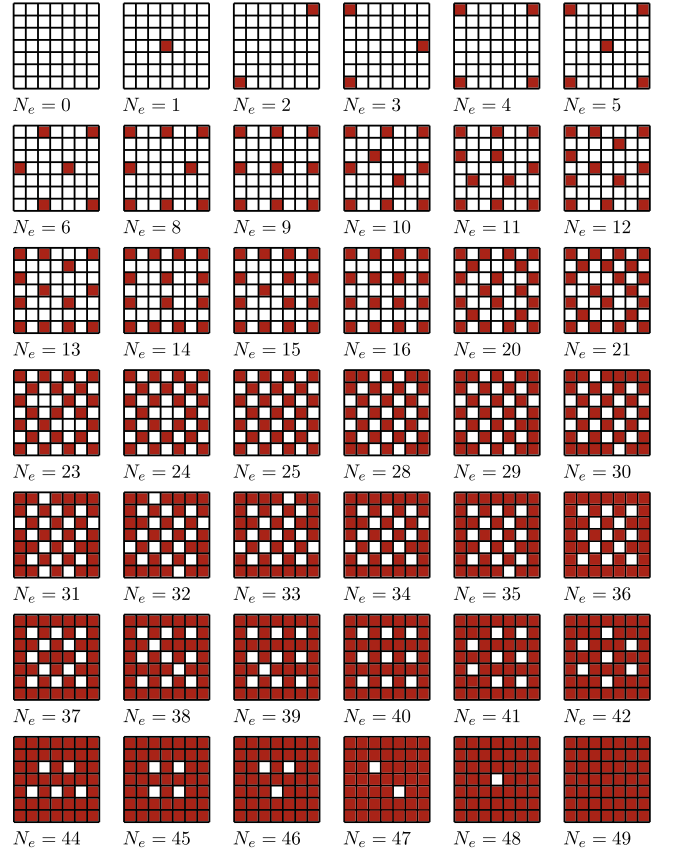}
\end{center}
\caption{(color online) Same as in \fig{\ref{fig:seq:square9}} but for $N=49$. The MECs for $N_e=7,17,18,19,22,26,27,43 $ are not shown since max-cut did not yield them as part of the ground state sequence up to a resolution in $\Delta/V_1$ of $10^{-8}$. Due to the computational complexity, we refrained from their determination by brute force methods as we did for smaller lattices.
}
\label{fig:seq:square49}
\end{figure}

In this paragraph we focus on the ground states of square lattices. The max-cut code allows us to determine the MECs for squares with up to 49 sites; the MECs are verified by the brute force method up to 36 sites only due to computation time. For the $7\times 7$ lattice, the resolution in $\Delta/V_1$ of max-cut was set to $10^{-8}$, \ie if the program misses MECs, they become a ground state for a range in $\Delta/V_1$ of less than $10^{-8}$. In Figs.\ \ref{fig:seq:square9}-\ref{fig:seq:square49} we give the resulting MECs and in Tables \ref{tab:seq:square9}-\ref{tab:seq:square49} we specify the range for which the MECs become the ground state. States which are ground state for a range of about $1V_1$ in $\Delta$ and more (`dominant ground states') are additionally emphasized by bold numbers in the tables.
Note that we again give only one representative MEC; the ground state in the symmetric subspace can be obtained by evaluating \eq{\ref{eq:mod:symmetrizing}} (except for $N_e=1$, where the ground state of the symmetric subspace is a superposition of symmetrized product states with $N_e=1$, \cf \sect{\ref{sec:gs:chain}}).

\begin{table}
\begin{ruledtabular}
	\begin{center}
		\caption{Square with $N=16$: Range for which the MECs in \fig{\ref{fig:seq:square16}} become a ground state.\label{tab:seq:square16}}
		\begin{tabular}{cc|cc|cc}
		$\Delta/V_1$ & $N_e$ & $\Delta/V_1$ & $N_e$ & $\Delta/V_1$ & $N_e$ \\
		\hline
		$-\infty$ & \textbf{16} & -3.042452 & \textbf{10} & -0.142953 & 4 \\
		-4.565203 & 15 & -2.019825 & 9 & -0.003003 & 3 \\
		-4.440203 & \textbf{14} & -2.019654 & \textbf{8} & -0.002655 & 2 \\
		-3.170405 & 13 & -0.533203 & 7 & -0.000171 & 1 \\
		-3.169405 & 12 & -0.411706 & 6 & 0 & \textbf{0} \\
		-3.043452 & 11 & -0.170292 & 5 & $\infty$ & \\
		-3.042452 &    & -0.142953 & & &
		\end{tabular}
	\end{center}
\end{ruledtabular}
\end{table}

The most interesting feature of the investigated 2D lattices is the possible lack of some $N_e$ in the sequence of ground states, \ie not every number of Rydberg atoms $N_e$ becomes a ground state at some value of $\Delta/V_1$. In the tables, we have emphasized such a lacking by underlining the next higher $N_e$; in the figures they are indicated by parentheses. A simple example is the $3\times3$ lattice. Its results are given in \fig{\ref{fig:seq:square9}} and \tab{\ref{tab:seq:square9}}.
The complete ground state sequence from positive to negative detuning contains the MECs with $N_e=0,1,2,3,4,5,8,9$ excitations. The MECs with $N_e=6,7$ never become a ground state. 

\begin{table}
\begin{ruledtabular}
	\begin{center}
		\caption{Square with $N=25$: Range for which the MECs in \fig{\ref{fig:seq:square25}} become a ground state.\label{tab:seq:square25}}
		\begin{tabular}{cc|cc|cc}
		$\Delta/V_1$ & $N_e$ & $\Delta/V_1$ & $N_e$ & $\Delta/V_1$ & $N_e$ \\
		\hline
		$-\infty$ & \textbf{25} & -3.066433 & 17 & -0.046457 & 6 \\
		-4.634313 & 24 & -3.066189 & \underline{16} & -0.032038 & 5 \\
		-4.556189 & 23 & -2.681863 & \textbf{13} & -0.007813 & 4 \\
		-4.526993 & 22 & -0.570313 & \underline{12} & -0.000543 & 3 \\
		-4.508212 & 21 & -0.504304 & 10 & -0.000464 & 2 \\
		-4.134313 & \textbf{20} & -0.504171 & 9 & -0.000031 & 1 \\
		-3.070340 & 19 & -0.071681 & 8 & 0 & \textbf{0} \\
		-3.070095 & 18 & -0.046625 & 7 & $\infty$ & \\
		-3.066433 &    & -0.046457 & & &
		\end{tabular}
	\end{center}
\end{ruledtabular}
\end{table}

To get a qualitative understanding of the omission of certain MECs, we consider the energy difference between two MECs $\ket{\alpha},\, \ket{\beta}$ with $N_e(\alpha)$ and $N_e(\beta)=N_e(\alpha)+n$ excitations ($n=1,2,\ldots$), respectively. Furthermore we assume that $\ket{\alpha}$ is a ground state for some detuning $\Delta'$. If we want to know for which laser detuning $\Delta$ the ground state changes from $\ket{\alpha}$ to $\ket{\beta}$ we need to fulfill
\begin{align}
 0>E(\beta) - E(\alpha) = n{\Delta} + \Eint(\beta) - \Eint(\alpha),
\label{eq:seq:energy_incr_full}
\end{align} 
according to \eq{\ref{eq:spec:evals_rescaled}}. We abbreviate the detuning where $E(\beta)= E(\alpha)$ by $\Delta_{N_e(\alpha),N_e(\beta)}$.
For our example, we set now $\ket{\alpha}$ to be the MEC with $N_e=5$ of the $3\times3$ lattice in \fig{\ref{fig:seq:square9}}, which has $\Eint(\alpha)/V_1\approx 0.6$. The MEC with $N_e=6$, denoted momentarily as $\ket{\beta_1}$, has $\Eint(\beta_1)/V_1\approx 3.6$, roughly $3V_1$ more than $\ket{\alpha}$.
Hence, $\ket{\beta_1}$ does only have a smaller energy eigenvalue than $\ket{\alpha}$ if $\Delta/V_1 \leq \Delta_{5,6}/V_1 \approx -3$. The same, $\Delta/V_1 \leq \Delta_{5,7}/V_1 \approx -2.9$, applies also to the MEC $\ket{\beta_2}$ with $N_e=7$ and $\Eint(\beta_2)/V_1\approx 6.4$. 
The MEC $\ket{\beta_3}$ with $N_e=8$ has in total $\Eint(\beta_3)/V_1\approx 8.7$, which means it reduces the energy if $\Delta/V_1 \leq \Delta_{5,8}/V_1 \approx -2.7$. 
Since the MEC $\ket{\beta_4}$ with $N_e=9$ has $\Eint(\beta_4)/V_1\approx 13.2$, its crossover with $\ket{\alpha}$ does only appear if $\Delta/V_1 \leq \Delta_{5,9}/V_1 \approx -3.2 $. Hence, when decreasing the detuning from $\Delta'$ on, the ground state sequence switches from the MEC with $N_e=5$ directly to the MEC with $N_e=8$ at $\Delta/V_1 = \Delta_{5,8}/V_1 \approx - 2.7 $, in agreement with \tab{\ref{tab:seq:square9}}. In other words, with respect to the MEC with $N_e=5$, the MEC with $N_e=8$ has the most beneficial ratio between additional interaction energy and additional excitations. This emphasizes once more that the ground state sequence and the crossover detunings crucially depend on the interaction energies of the MECs in the considered lattice.

\begin{table}
\begin{ruledtabular}
	\begin{center}
		\caption{Square with $N=36$: Range for which the MECs in \fig{\ref{fig:seq:square36}} become a ground state.\label{tab:seq:square36}}
		\begin{tabular}{cc|cc|cc}
		$\Delta/V_1$ & $N_e$ & $\Delta/V_1$ & $N_e$ & $\Delta/V_1$ & $N_e$ \\
		\hline
		$-\infty$ & \textbf{36} & -3.053621 & 23 & -0.145741 & 9 \\
		-4.643048 & 35 & -3.053438 & 22 & -0.033635 & 8 \\
		-4.600027 & 34 & -3.028838 & 21 & -0.030593 & 7 \\
		-4.584402 & 33 & -3.028813 & \textbf{20} & -0.018665 & 6 \\
		-4.570631 & \underline{32} & -2.020436 & 19 & -0.006522 & 5 \\
		-4.433069 & 30 & -2.020428 & \textbf{18} & -0.003035 & 4 \\
		-4.229609 & 29 & -0.574484 & \underline{17} & -0.000142 & 3 \\
		-4.109845 & \textbf{28} & -0.542782 & 15 & -0.000122 & 2 \\
		-3.170676 & 27 & -0.537173 & \underline{14} & -0.000008 & 1 \\
		-3.170650 & 26 & -0.394663 & 12 & 0 & \textbf{0} \\
		-3.070303 & 25 & -0.179229 & 11 & $\infty$ & \\
		-3.070120 & 24 & -0.164336 & 10 &  & \\
		-3.053621 &    & -0.145741 & & & 
		\end{tabular}
	\end{center}
\end{ruledtabular}
\end{table}

A similar situation also appears for the other considered square lattices with an odd number of sites ($N=25$ and $49$, Figs.\ \ref{fig:seq:square25} and \ref{fig:seq:square49}) for the MEC with $N_e=\lceil N/2 \rceil$. 
Qualitatively, this can be explained as follows: In all these observed cases the MEC with $N_e=\lceil N/2 \rceil$ has a checkerboard-like structure with excited corners, \cf \fig{\ref{fig:seq:square25}}, $N_e=13$. Adding further excitations to this excitation pattern is unfavorable since it increases the interaction energy by at least 3 neighboring excitations. However, if one rearranges the excitations and forms the inverted structure it is possible to add excitations at the corners, which increases the interaction energy only by 2 neighboring excitations. As a result it turns out that the MECs with $N_e=\lceil N/2 \rceil+1,\,\lceil N/2 \rceil+2$ are omitted and the ground state sequence has a crossover between $N_e=\lceil N/2 \rceil$ and $N_e=\lceil N/2 \rceil+3$. 
The lattices show additional omissions of single MECs, but this depends on the individual case of the lattice.

Interestingly, the $4\times4$ lattice is not missing any MEC. This cannot be generalized to all square lattices with an even number of sites as the case of the $6\times6$ lattice demonstrates, \cf \fig{\ref{fig:seq:square36}}.

\begin{table}
\begin{ruledtabular}
	\begin{center}
		\caption{Square with $N=49$: Range for which the MECs in \fig{\ref{fig:seq:square49}} become a ground state. \label{tab:seq:square49}}
		\begin{tabular}{cc|cc|cc}
		$\Delta/V_1$ & $N_e$ & $\Delta/V_1$ & $N_e$ & $\Delta/V_1$ & $N_e$ \\
		\hline
		$-\infty$ & \textbf{49} & -3.070831 & 34 & -0.071157 & 14 \\
		-4.652127 & 48 & -3.070685 & 33 & -0.059398 & 13 \\
		-4.635938 & 47 & -3.070492 & 32 & -0.057176 & 12 \\
		-4.618859 & 46 & -3.053305 & 31 & -0.047315 & 11 \\
		-4.606706 & 45 & -3.053112 & 30 & -0.041627 & 10 \\
		-4.563284 & \underline{44} & -3.052966 & 29 & -0.033062 & 9 \\
		-4.553749 & 42 & -3.052950 & \underline{28} & -0.006277 & \underline{8} \\
		-4.506435 & 41 & -2.681913 & \textbf{25} & -0.004117 & 6 \\
		-4.506287 & 40 & -0.578998 & 24 & -0.002765 & 5 \\
		-4.140838 & 39 & -0.570054 & \underline{23} & -0.000686 & 4 \\
		-4.138884 & 38 & -0.546426 & 21 & -0.000048 & 3 \\
		-4.135408 & 37 & -0.541998 & \underline{20} & -0.000041 & 2 \\
		-4.081814 & \textbf{36} & -0.493574 & 16 & -0.000003 & 1 \\
		-3.070846 & 35 & -0.089938 & 15 & 0 & \textbf{0} \\
		-3.070831 &    & -0.071157 &    & $\infty$ & 
		\end{tabular}
	\end{center}
\end{ruledtabular}
\end{table}

\subsubsection{Rectangular lattices}

We end the discussion of ground state sequences with a brief outlook on rectangular lattices. Depending on the geometry, we were able to evaluate the ground states of rectangular lattices with up to 48 sites by means of max-cut. 

\begin{figure}
\begin{center}
\includegraphics[width=8cm]{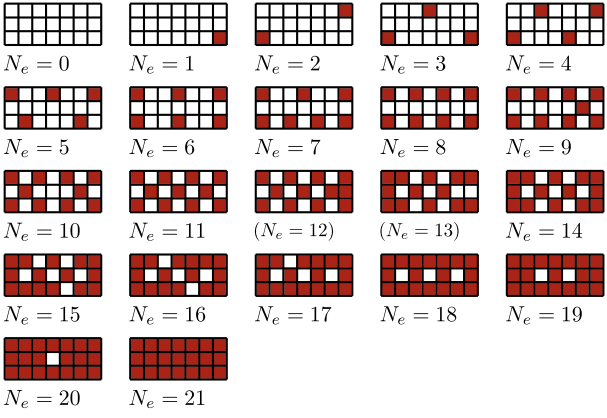}
\end{center}
\caption{(color online) Same as in \fig{\ref{fig:seq:square9}} but for a rectangle $N=3\times 7$.}
\label{fig:seq:rectangle_21_7}
\end{figure}
\begin{table}
\begin{ruledtabular}
	\begin{center}
		\caption{Rectangle with $N=21= 3\times 7$: Range for which the MECs in \fig{\ref{fig:seq:rectangle_21_7}} become a ground state.\label{tab:seq:rectangle21-7}}
		\begin{tabular}{cc|cc|cc}
		$\Delta/V_1$ & $N_e$ & $\Delta/V_1$ & $N_e$ & $\Delta/V_1$ & $N_e$ \\
		\hline
		$-\infty$ & \textbf{21} & -3.051978 & \underline{14} & -0.031259 & 5 \\ 
		-4.569993 & 20 & -2.689414 & \textbf{11} & -0.016634 & 4 \\ 
		-4.548927 & 19 & -0.535250 & 10 & -0.005432 & 3 \\ 
		-4.491230 & \textbf{18} & -0.502358 & 9 & -0.000916 & 2 \\ 
		-3.085181 & 17 & -0.502114 & 8 & -0.000016 & 1 \\ 
		-3.083228 & 16 & -0.067022 & 7 & 0 & \textbf{0} \\
		-3.053931 & 15 & -0.048241 & 6 & $\infty$ & \\
		-3.051978 &    & -0.031259 & & & 
		\end{tabular}
	\end{center}
\end{ruledtabular}
\end{table}

Due to the limited scope of this work we restrict ourselves to present only two examples: $N=21=3\times 7$ and $N=24=3\times8$. 
In the first case we have an odd number of sites. The results are summarized in \fig{\ref{fig:seq:rectangle_21_7}} and \tab{\ref{tab:seq:rectangle21-7}}. Similar to square lattices with an odd number of sites we find that the MECs with $N_e=12$ and $N_e=13$ are omitted in the ground state sequence. The explanation is analogous to the discussion of the $3\times3$ lattice: The MEC with $N_e=11$ forms a checkerboard-like pattern while the excitations in the MEC with $N_e=14$ resembles the inverted pattern with excited corners. The associated interaction energies lead to a omission of the intermediate MECs. We remark that also the other investigated rectangular lattice with 3 rows and an odd number of sites systematically skip the MECs with $N_e=\lceil \frac{N}{2}\rceil + 1$ and $N_e=\lceil \frac{N}{2}\rceil + 2$, since they undergo similar arrangements of the excitations as the $3\times7$ lattice.

\begin{figure}
\begin{center}
\includegraphics[width=8.5cm]{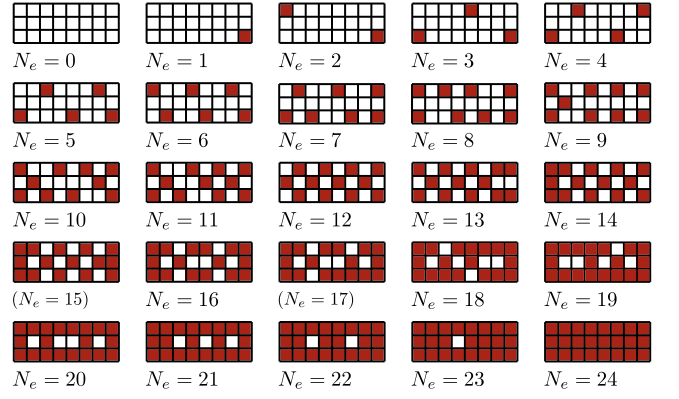}
\end{center}
\caption{(color online) Same as in \fig{\ref{fig:seq:square9}} but for a rectangle $N=3\times 8$.}
\label{fig:seq:rectangle_24_8}
\end{figure}
\begin{table}
\begin{ruledtabular}
	\begin{center}
		\caption{Rectangle with $N=24= 3\times 8$: Range for which the MECs in \fig{\ref{fig:seq:rectangle_24_8}} become a ground state.\label{tab:seq:rectangle24-8}}
		\begin{tabular}{cc|cc|cc}
		$\Delta/V_1$ & $N_e$ & $\Delta/V_1$ & $N_e$ & $\Delta/V_1$ & $N_e$ \\
		\hline
		$-\infty$ & \textbf{24} & -2.901020 & 14 & -0.046089 & 6 \\
		-4.570645 & 23 & -2.268020 & 13 & -0.017949 & 5 \\
		-4.562885 & 22 & -1.783945 & \textbf{12} & -0.011288 & 4 \\
		-4.508958 & \textbf{21} & -0.560548 & 11 & -0.003907 & 3 \\
		-3.536517 & 20 & -0.521419 & 10 & -0.000582 & 2 \\
		-3.083073 & 19 & -0.491402 & 9 & -0.000007 & 1 \\
		-3.078055 & \underline{18} & -0.277148 & 8 & 0 & \textbf{0} \\
		-3.058263 & \underline{16} & -0.046921 & 7 & $\infty$ & \\
		-2.901020 &    & -0.046089 & & & 
		\end{tabular}
	\end{center}
\end{ruledtabular}
\end{table}

In the case of 24 sites we observe that the MECs with $N_e=15$ and $N_e=17$ are omitted, \cf \fig{\ref{fig:seq:rectangle_24_8}} and \tab{\ref{tab:seq:rectangle24-8}}. It turns out that the MEC $N_e=\lceil \frac{N}{2}+3\rceil$ is also skipped in the other investigated rectangular lattices with 3 rows and an even number of sites. The reason for this are arrangements of the excitations comparable to those of the lattice with 24 sites. 
In summary, seemingly the excitation patterns of the MECs of all rectangular lattices with 3 rows follow comparable arrangements of the excitations with increasing $N_e$. This might be due to the small size of the lattice.

\section{Summary and Outlook}
\label{sec:summary}

We explored the symmetries and spectral properties of two-dimensional, laser-driven lattices with a main focus on square lattices. Exploiting the geometry induced symmetries allowed us to truncate the state space when initializing the system in the canonical ground state $\ket{G}=\ket{gg\ldots g}$, which facilitates the numerical treatment. 

An investigation of the spectra in the limit of vanishing laser coupling of square lattices showed that the spectra of small square lattices show similar features as the linear chain. Specifically, we observe in the spectra also points/regions of high degeneracy and large energetic gaps between them. However these features appear on a scale reduced by a factor of 8 which is due to the diagonal-neighbor contributions. Because of the corrections originating from long-range contributions, these properties are less pronounced with increasing lattice size.

For the investigation of the laser detuning-dependent ground states, we have demonstrated that in the limit of vanishing laser coupling the model Hamiltonian can be written in the form of an Ising-Hamiltonian with a site-dependent external magnetic field. The ground states of the obtained Ising-Hamiltonian have been evaluated for various one- and two-dimensional lattices by using elements of graph theory. The method used in this work, however, is in principle applicable to arbitrary lattice geometries.
In contrast to the case of the linear chain, where there is a state for every number of excitations $N_e$ which becomes a ground state at some laser detuning, we observed that this is in general not the case for two-dimensional lattices. We emphasize that the results presented here cannot be implied to occur in a similar way for larger lattices or other interaction potentials since the optimization of the interaction energy crucially depends on the lattice and interaction potential.

As indicated in Sec.\ \ref{sec:ground_states}, our restriction to a vanishing laser coupling limits the validity of the presented ground states sequences. While we expect the dominant ground states still to be promising ground state candidates in presence of finite laser couplings, it is \emph{a priori} not clear to what extend this holds, especially for nearly degenerate ground states and for the crossover regimes. Answering this question is subject of future inquiries and necessitates an even higher computational effort due to the requirement of numerically diagonalizing the general Hamiltonian (\ref{eq:mod:gen_ham}).

Another possibly diminishing effect on the predictive power of the ground state sequences presented in the present work is the presence of disorder and defects in the Rydberg lattice. We expect that defects can have a quite dramatic role on the actual ground state geometry, for example, when a defect in the middle of a chain effectively splits it into two separate ones. Disorder, on the other hand, we believe to act more subtle. Especially energetically separated dominant ground states should be quite robust against a small jitter in the lattice constant. But again, only a comprehensive future study can tell us more.

\begin{acknowledgments}
M.M. acknowledges financial support by a fellowship within the postdoc-programme of the German Academic Exchange Service (DAAD).
P.S. acknowledges financial support by the Initial Training Network COHERENCE (Marie-Curie Actions) of the program FP7 of the European Union.
\end{acknowledgments}

\begin{appendix}
\section{Rewriting the Hamiltonian diagonal\label{sec:app:appendix}}
In this appendix we demonstrate that \eq{\ref{eq:gs:diag_ham}} can be rewritten by only using $\sigma^{(k)}_z$-matrices as in \eq{\ref{eq:gs:diag_sigmaz}}.

For the interaction term $H_{\text{int}}=\sum^{N-1}_{k=1}\sum^N_{j=k+1} V_{k,j} n^{(k)}_e n^{(j)}_e$ in \eq{\ref{eq:gs:diag_ham}} we use $n^{(k)}_e = \frac12 [\sigma^{(k)}_z + \mathbbm{1}]$. Neglecting momentarily the occurring constant energy offset $C=\frac14 \sum^{N-1}_{k=1}\sum^N_{j=k+1} V_{k,j}$, the interaction term $H_{\text{int}}$ reads
\begin{align}
\frac14 \sum^{N-1}_{k=1}\sum^N_{j=k+1} V_{k,j} \sigma^{(k)}_z \sigma^{(j)}_z + \frac14 \sum^{N-1}_{k=1}\sum^N_{j=k+1} V_{k,j} [\sigma^{(k)}_z + \sigma^{(j)}_z ].
\end{align}

The first term has already the aspired form and we rearrange the individual parts of the sum of the second part with respect to $\sigma^{(k)}_z$. When defining a number $r_k$ such that
\begin{align}
 \sum^{N-1}_{k=1}\sum^N_{j=k+1} V_{k,j} \left(\sigma^{(k)}_z + \sigma^{(j)}_z \right) =: \sum^N_{k=1} r_k \sigma^{(k)}_z,
\label{eq:app:rewriting_linear}
\end{align}
one finds
\begin{align}
 r_k = \sum^N_{j=k+1} V_{k,j} + \sum^{k-1}_{j=1} V_{j,k} =  \sum^N_{\substack{j=1\\j\neq k}} V_{k,j}, \qquad k=1\ldots N.
\label{eq:app:rk}
\end{align}
Please note, the last equality holds only if the interaction potential satisfies $V_{k,j}=V_{j,k}$, as is the case for our system.
We demonstrate briefly the equality in \eq{\ref{eq:app:rewriting_linear}} by inserting \eq{\ref{eq:app:rk}}:
\begin{align}
 \sum^N_{k=1} r_k \sigma^{(k)}_z &= \sum^N_{k=1} \left( \sum^N_{j=k+1} V_{k,j} + \sum^{k-1}_{j=1} V_{j,k} \right) \sigma^{(k)}_z\\
  &= \sum^{N-1}_{k=1} \sum^N_{j=k+1} V_{k,j} \sigma^{(k)}_z + \sum^N_{k=2} \sum^{k-1}_{j=1} V_{j,k} \sigma^{(k)}_z.
\end{align}
Note, we dropped those indices in the last line where empty sums occur. 
The first term is as desired and we remain with showing 
\begin{align}
 \sum^{N-1}_{k=1}\sum^N_{j=k+1} V_{k,j} \sigma^{(j)}_z = \sum^N_{k=2} \sum^{k-1}_{j=1} V_{j,k} \sigma^{(k)}_z,
\label{eq:app:to_prove_induction}
\end{align}
which can be done by employing induction. Starting with $N=2$, \eq{\ref{eq:app:to_prove_induction}} becomes
\begin{align}
  V_{1,2} \sigma^{(2)}_z &= V_{1,2} \sigma^{(2)}_z.
\end{align}
Under the presumption that \eq{\ref{eq:app:to_prove_induction}} holds for $N$ sites we can show that it also holds for $N+1$ sites:
\begin{align}
 \sum^{(N+1)-1}_{k=1}&\sum^{N+1}_{j=k+1} V_{k,j} \sigma^{(j)}_z= \nonumber\\
&=\sum^{N}_{k=1}\left(\sum^{N}_{j=k+1} V_{k,j} \sigma^{(j)}_z + V_{k,N+1} \sigma^{(N+1)}_z \right)\nonumber\\
&=\sum^{N-1}_{k=1}\sum^{N}_{j=k+1} V_{k,j} \sigma^{(j)}_z + \sum^{N}_{k=1}  V_{k,N+1} \sigma^{(N+1)}_z\nonumber\\
&\stackrel{(\ref{eq:app:to_prove_induction})}{=} \sum^{N}_{k=2} \sum^{k-1}_{j=1} V_{j,k} \sigma^{(k)}_z + \sum^{N}_{j=1} V_{j,N+1} \sigma^{(N+1)}_z\nonumber\\
&=\sum^{N+1}_{k=2} \sum^{k-1}_{j=1} V_{j,k} \sigma^{(k)}_z.
\end{align}
Thus, recalling the constant energy offset $C$, in summary one can write $H_{\text{int}}$ as follows
\begin{align}
 H_{\text{int}} ={}& \sum_{k=1}^{N} \frac14 \left(\sum_{j\neq k} V_{k,j}\right)  \sigma^{(k)}_z \nonumber\\
&+ \frac14 \sum_{k=1}^{N-1} \sum_{j=k+1}^{N} V_{k,j} \sigma^{(k)}_z \sigma^{(j)}_z+C.
\end{align}

Together with the laser detuning part $H_{\text{det}}=\frac{\Delta}{2} \sum_{k=1}^{N} \sigma^{(k)}_z$, the diagonal Hamiltonian in \eq{\ref{eq:gs:diag_ham}} is equivalent to
\begin{align}
 H_{\text{diag}} ={}& \sum_{k=1}^{N} \left( \frac{\Delta}{2} + \frac14 \sum_{j\neq k} V_{k,j} \right)  \sigma^{(k)}_z \nonumber\\
 &+\frac14 \sum_{k=1}^{N-1} \sum_{j=k+1}^{N} V_{k,j} \sigma^{(k)}_z \sigma^{(j)}_z+C.
\end{align}
\end{appendix}

\bibliography{paper}

\end{document}